\documentclass[amsmath,amssymb,prd,twocolumn,nofootinbib,floatfix]{revtex4}

\usepackage{graphicx}
\usepackage{epsf}
\usepackage[title]{appendix}
\usepackage[latin2]{inputenc}

\newcommand{\Mpc}{{\ensuremath{\rm Mpc}}}

\newcommand{\vk}{\mathbf k}
\newcommand{\vp}{\mathbf p}
\newcommand{\vq}{\mathbf q}

\newcommand{\vx}{\mathbf x}
\newcommand{\vu}{\mathbf u}
\newcommand{\vL}{\mathbf L}
\newcommand{\vpsi}{\mathbf \Psi}

\newcommand{\vxi}{\mathbf{\Xi}}

\newcommand{\mF}{\mathcal{F}}
\newcommand{\mK}{\mathcal{K}}
\newcommand{\mH}{\mathcal{H}}

\newcommand{\Rmnum}[1]{\uppercase\expandafter{\romannumeral #1}}

\newcommand{\psid}{\Xi_{m}}

\newcommand{\pr}{\prime}

\newcommand{\vkpr}{\vk^{\pr}}
\newcommand{\vppr}{\vp^{\pr}}

\newcommand{\zobs}{z_{\it obs}}
\newcommand{\zform}{z_{\it form}}

\begin{document}
\title{Resummed Perturbation Theory of Galaxy Clustering}
\author{Xin Wang}
\email{wangxin@pha.jhu.edu}
\author{Alex Szalay}
\affiliation{Department of Physics and Astronomy, Johns Hopkins University, Baltimore, MD 21218, US}
\date{\today}


\begin{abstract}
    The relationship between observed tracers such as galaxies and the underlying 
    dark matter distribution is crucial in extracting cosmological information.
    As the linear bias model breaks down at quasi-linear scales, the standard perturbative
    approach of the nonlinear Eulerian bias model (EBM) is not accurate enough in 
    describing galaxy clustering. In this paper, we discuss such a model in the 
    context of resummed perturbation theory, and further generalize it to incorporate 
    the subsequent gravitational evolution by combining with a Lagrangian 
    description of galaxies' motion.
    The multipoint propagators we constructed for such model also exhibit 
    exponential damping similar to their dark matter counterparts, 
    therefore the convergence property of statistics built upon these quantities 
    is improved. This is achieved by applying both Eulerian and Lagrangian resummation
    techniques of dark matter field developed in recent years.
    As inherited from the Lagrangian description of galaxy density evolution, our approach 
    automatically incorporates the non-locality induced by gravitational evolution 
    after the formation of the tracer, and also allows us to include a continuous 
    galaxy formation history by temporally weighted-averaging relevant quantities with 
    the galaxy formation rate.
\end{abstract}
\maketitle

\section{Introduction}

Large-scale structure surveys provide a wealth of cosmological information in 
probing dark energy, modified gravity, neutrino masses and the  
physics of early universe. 
As the statistical uncertainty decreases dramatically in next generation surveys, 
it requires us to achieve an unprecedented level of accuracy in theoretically predicting 
the statistics of the observed clustering pattern. 
Since nonlinear dynamics plays a important role in understanding the observed data
in relevant regime, it has attracted lots of interest in developing 
the nonlinear perturbation theory of the dark matter beyond the standard approach.
Many sophisticated methods have been proposed, including  the renormalized 
perturbation theory (RPT) \citep{CS06a,CS06b,BCS08}, the Lagrangian resummation 
theory \citep{M08a}, the closure theory \citep{TH08},  and the time renormalization 
group theory (TRG) \citep{P08}. 
However, the relationship between various observable tracers (galaxies, quasars,
Ly$\alpha$ forest, clusters and HI galaxies etc.) and the underlying dark matter field
 may be more challenging.
The linear bias model breaks down even for present surveys \citep{SC08,P07} at relatively
large scale, the scale-dependence observed in the data varies among different galaxy
samples as one would expect \citep{SC08}, and the empirical fitting formula \citep{C05} 
widely used in the data analysis is not able to describe such differences among samples
and eventually leads to inconsistent cosmological constraints \citep{SC08}.

Two different approaches exist in modeling the scale-dependent bias of 
galaxies these days.
One is the halo-based model which starts from the large scale clustering of dark
matter halos and then populates with different types of galaxies according to the 
halo occupation distribution (HOD) function.
The other approach is the perturbative bias model, which packs the complicated
baryonic physics into a general unknown functional depending on the dark matter
distribution, and characterize it with the first few bias parameters.
With a careful parameterized and calibrated HOD function, the first approach can 
describe the two-point statistics qualitatively well at small scale. 
However, a viable HOD parameterization usually contains lots of free parameters, 
and heavily depends on calibration from simulation.
Moreover, in the quasi-linear regime where most of large-scale surveys probe, 
the accuracy of such approach is usually not satisfactory.

On the other hand, the standard perturbative bias model is also problematic.
In the Eulerian bias model (EBM), the observed galaxy number density depends on the 
Eulerian dark matter distribution at the time of observation. 
Therefore, this model breaks down as the dark 
matter perturbations evolve into the nonlinear regime at small scales and later times.
Since both the large-scale amplitude and small-scale shape of the power spectrum depend
on higher-order bias parameters, such calculation would fail within the whole range of 
scales.
Furthermore, although adopting some artificial nonlocal bias parameters won't bring too
 much intrinsic difficulties in the calculation, most authors simply assume that the
  functional form of the galaxy density only depends on the matter distribution locally, 
which is not consistent with the nonlocal gravitational evolution of the 
tracer \citep{M11,CSS12}.

Since the difficulties of the perturbative convergence encountered in the nonlinear 
bias model is very similar to the situation in dark matter field, one would expect 
to generalize any of 
previous nonstandard matter perturbation theory to describe the clustering of biased
tracer. Several works have already been done along this direction. 
In \cite{M08b}, a Lagrangian bias model (LBM) which intended to incorporate the halo 
bias and redshift distortion was proposed based on a similar resummation theory for
matter distribution \citep{M08a}.
Then \cite{EK11} also included the halo bias in the time renormalization group methods.
\cite{Mc06,Mc09} approached in a different way by redefining the bias parameters. 
As we will show in this paper, with the help of recent development in the renormalized
perturbation theory \cite{CS06a,CS06b,BCS08}, it is also possible to construct a resummed
perturbation theory for nonlinear Eulerian bias model, and the perturbative expansion 
of statistics can be rewritten through the so-called $\Gamma-$expansion, first 
introduced by \cite{BCS08}, with the building blocks known as multipoint propagators.
One advantage of the $\Gamma-$expansion is that for Gaussian initial conditions,
all contributions are positive and centralized within a limited
range of scale. As we will see later, such an expansion is quite general and 
applicable to many circumstances as long as the multipoint propagators can be well
estimated. Therefore in the following, this paper will mainly concentrate on
 constructing the multipoint propagators for various objects.
The validity of such technique has already been verified in calculating the power 
spectrum of logarithmic mapping of the density field \cite{W11}. 
It have been shown that even for the slow converging function such as logarithm, 
the calculation agrees with the simulation quite well.

We further generalize the Eulerian bias model by separating the complicated
nonlinear physical processes during the galaxy formation and the subsequent
gravitational evolution until been observed by a survey. 
As we will show, this can be achieved by combining the Eulerian bias model and 
Lagrangian description of galaxy motion introduced in \cite{M08b}. 
Like EBM, we assume the galaxy distribution created at some time 
is characterized by a general unknown functional, and the galaxy number density observed 
at a Eulerian position later can be naturally described by the Lagrangian version of continuity equation. 
Consequently, this approach also enables us to incorporate continuous galaxy formation
by temporally averaging the distribution of newborn galaxies weighted with the galaxy 
formation rate.
In such a framework, while the complex galaxy formation physics is 
simplified by a few bias parameters, we are able to capture the following 
gravitational evolution after its creation, which would in principal induce non-locality
 and can not be accurately described by a local Eulerian bias model.
As inherited from the Lagrangian bias model and our resumed calculation of 
Eulerian bias model, this model can also be expressed in a resummed manner.

In section II, we briefly review both Eulerian and Lagrangian perturbation
theory of dark matter field, including both standard approach and resummed theory.
We will also introduce the concept of the multipoint propagator in the context of 
Lagrangian perturbation theory, and give a explicit calculation in Appendix A.
In section III, we first review the standard Eulerian and Lagrangian bias model, 
and then introduce our continuous galaxy formation model. 
We construct the resummed Eulerian bias model in Section IV, and the generalize
it in Section V. Finally, we conclude in Section VI.

\section{Perturbation Theory of Dark Matter}
In this section, we will briefly review both Eulerian perturbation theory
(EPT) and the Lagrangian perturbation theory (LPT) of the dark matter field, including
both the standard approach and resummed calculation.

\subsection{Dynamical Equation of Motion}
The gravitational dynamics of a pressureless fluid before shell
crossing is governed by the continuity, Euler, and Poisson
equations.
\begin{eqnarray}
\label{eqn:dyneqn_sep}
 \frac{\partial \delta(\vx, \tau)}{\partial \tau} + \nabla \cdot 
\bigl [~ \bigl (1+\delta(\vx,\tau) \bigr) ~ \vu(\vx, \tau) ~\bigr ] 
= 0,\qquad\quad \nonumber \\
\frac{\partial \vu(\vx,\tau)}{\partial \tau} + \mH(\tau) \vu(\vx, \tau)
+\vu(\vx, \tau) \cdot \nabla\vu(\vx, \tau) \qquad\quad  \nonumber \\  
 = -\nabla \Phi_N(\vx, \tau)  \nonumber\\
\nabla^2 \Phi_N(\vx, \tau) = \frac{3}{2} \Omega_m(\tau) \mH^2(\tau) 
\delta(\vx, \tau). \qquad\qquad\qquad
\end{eqnarray}
Here, $\mH= d\ln a(\tau)/ d\ln\tau$ is the Hubble expansion rate, $a(\tau)$ is
the scale factor,
$\Omega_m(\tau)$ is the ratio of matter density to critical density,
and $\Phi_N(\vx,\tau)$ is the Newtonian potential.

Following \cite{CS06a,CS06b}, the equation of motion in Fourier space can be 
written in a compact form by defining the two-component variable
\begin{eqnarray}
\vxi_a(\vk, \eta) = \bigl( \delta(\vk, \eta), ~ -\theta(\vk, \eta)/\mH \bigr), 
\end{eqnarray}
where $\theta$ denotes the divergence of peculiar velocity $\nabla \cdot \vu$.
The index $a\in\{1, 2\}$ stands for matter density or velocity variables, respectively.
In the following, we will interchangeably use $\Xi_1$, $\Xi_m$ and
$\delta_m$ for matter overdensity, while using both $\Xi_g$ and $\delta_g$ for 
galaxy perturbation.

The equation of motion then reads
\begin{eqnarray}
 \label{eqn:dyneqn}
 \partial_{\eta} \vxi_a(\vk, \eta) &+ &\Omega_{ab} \vxi_b(\vk,\eta)  \nonumber\\
  &=& \gamma_{abc}
 (\vk,\vk_1,\vk_2) \vxi_b(\vk_1, \eta) \vxi_c(\vk_2, \eta),
\end{eqnarray}
with the convention that repeated Fourier arguments are integrated
over.  The time $\eta=\ln a(\tau)$ in a Einstein-de Sitter (EdS)
universe.  Here the constant matrix
\begin{eqnarray}
 \Omega_{ab} = \left[ \begin{array}{cc} 0 & -1\\ -3/2~ & 1/2 \end{array} \right ],
\end{eqnarray}
derived for an EdS universe, is still applicable in other cosmologies
with negligible corrections to the coefficients, using $\eta=\ln
D(\tau)$ (with $D$ the linear growth factor).

The symmetrized vertex matrix $\gamma_{abc}$ is given by
\begin{eqnarray}
\gamma_{222}(\vk,\vk_1,\vk_2) &=& \delta_D(\vk-\vk_1-\vk_2) 
\frac{|\vk_1+\vk_2|^2 (\vk_1 \cdot \vk_2)}{2 k_1^2 k_2^2} \nonumber \\
\gamma_{121}(\vk,\vk_1,\vk_2) &=& \delta_D(\vk-\vk_1-\vk_2) 
\frac{(\vk_1+\vk_2) \cdot \vk_1 }{2 k_1^2 }
\end{eqnarray}
$\gamma_{112}(\vk,\vk_1,\vk_2)=\gamma_{121}(\vk,\vk_2,\vk_1)$, and $\gamma=0$
otherwise.

Then the formal integral solution to Eq.\ (\ref{eqn:dyneqn}) can be derived as
\begin{eqnarray}
\label{eqn:solution}
\vxi_a(\vk,\eta) &=& g_{ab}(\eta) \phi_b(\vk) + 
\int_0^{\eta} d\eta^{\pr} ~ g_{ab}(\eta-\eta^{\pr}) \nonumber \\
  && \times  \gamma_{bcd}(\vk, \vk_1, \vk_2) 
  \vxi_c(\vk_1, \eta^{\pr}) 
 \vxi_d(\vk_2, \eta^{\pr})
\end{eqnarray}
where $\phi_a(\vk)$ denotes the initial condition 
$\phi_a(\vk)\equiv \Xi_a(\vk,\eta=0)$, and the linear propagator $g_{ab}(\eta)$
is given by
\begin{eqnarray}
 g_{ab}(\eta) = \frac{e^{\eta}}{5} \left[ \begin{array}{cc} 3 & 2\\3 & 2 
 \end{array} \right ] - \frac{e^{-3\eta/2}}{5} 
 \left[ \begin{array}{cc} -2 & 2\\3 & -3 \end{array} \right ].
\end{eqnarray}
In the following, we adopt growing-mode initial conditions 
$\phi_a(\vk) = \delta_0(\vk) u_a $, with $u_a=(1, 1)$.

In Lagrangian perspective, the dynamic of a fluid element at initial Lagrangian 
position $\vq$ is entirely described by the final Eulerian position $\vx$,
or equivalently the displacement field $\vpsi(\vq)= \vx-\vq$, which
is governed by the equation of motion,
\begin{eqnarray}
    \label{eqn:LPT_dynamics}
    \frac{d^2}{d\tau^2} \vpsi + \mH(\tau) \frac{d}{d\tau} \vpsi = -\nabla \Phi_N
\end{eqnarray}
where $\nabla$ is still the derivative with respect to Eulerian coordinates $\vx$,  and
the gravitational potential $\Phi_N$ is determined by Poisson equation as shown 
in Eq. (\ref{eqn:dyneqn_sep}). The density contrast is then related to $\vpsi(\vq)$
via mass conservation
\begin{eqnarray}
    \label{eqn:mass_conserv}
    1+\delta_m(\vx) = J^{-1} = \biggl [ {\rm det}\left [\delta_{ij}+ \Psi_{i,j}(\vq) 
    \right] \biggr]^{-1}
\end{eqnarray} 
where $J$ is the Jacobian of the transformation between Eulerian and Lagrangian 
space, and $\Psi_{i,j}= \partial \Psi_i / \partial q_j$. 
To linear order, Eq. (\ref{eqn:LPT_dynamics}) is then solved by the Zel'dovich 
approximation $\nabla_q \cdot \vpsi^{(1)}(\vq)= - D(\tau) \delta_0(\vq)$, with 
Lagrangian derivative $\nabla_q$.

\subsection{Eulerian Perturbation Theory}
A perturbative solution to Eq.\ (\ref{eqn:solution}) could be obtained by expanding
in terms of initial fields
\begin{eqnarray}
\label{eqn:psid}
 \vxi_a(\vk, \eta) &=& \sum_{n=1}^{\infty} \vxi^{(n)}_a (\vk, \eta) \nonumber \\
 \vxi^{(n)}_a(\vk, \eta) 
   &=& \int d^3\vq_{1 \cdots n} ~\delta_D(\vk-\vq_{1\cdots n})
  \mF^{(n)}_{a b_1 \cdots b_n}(\vq_1, \cdots, \nonumber \\ &&  \cdots, \vq_n;\eta) 
    \phi_{b_1}(\vq_1) \cdots \phi_{b_n}(\vq_n) 
\end{eqnarray}
where $d^3\vq_{1 \cdots n}$ is short for $d^3\vq_1 \cdots d^3\vq_n $, and
$\vq_{1\cdots n}$ denotes $\vq_1+\cdots+\vq_n$. The kernels $\mF^{(n)}$ are fully 
symmetric functions of the wave vectors. As shown in \cite{CS06b}, they can be 
obtained in terms of $g_{ab}$ and $\gamma_{abc}$ recursively.
\begin{eqnarray}
&\mF^{(n)}_a(\vk_1,\cdots,\vk_n; \eta) \delta_D(\vk-\vk_{1\cdots n})= \qquad\qquad 
   \nonumber \\
& \quad \biggl[ \sum_{m=1}^n\int_0^{\eta} d\eta^{\pr}  g_{ab}(\eta-\eta^{\pr})
~ \gamma_{bcd}(\vk, \vk_{1\cdots m}, \vk_{m+1 \cdots n}) \nonumber \\
& \qquad \times~ \mF_c^{(m)} (\vk_{1\cdots m}; \eta^{\pr}) 
 ~\mF_c^{(n-m)} (\vk_{m+1\cdots n}; \eta^{\pr}) \biggr]_{\rm symmetrized}
\end{eqnarray}
For $n=1$, $\mF^{(1)}_a(\eta)= g_{ab}(\eta) u_b$.
It should be noted that in this formalism, the kernel depends on the 
time $\eta$, since it includes subleading terms in $e^{\eta}$ \citep{CS06a}. 
If one only considers the fastest-growing mode, $\mF^{(n)}_a(\eta)$ equals the well-known PT 
kernel $D^n(\eta) \{F^{(n)}, G^{(n)}\}$, where $D(\eta)$ is the linear growth factor.

\begin{figure}[htp]
\begin{center}
\includegraphics[width=0.45\textwidth]{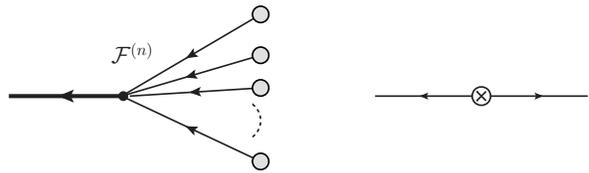}
\end{center}
\caption{\label{fig:EPT_mat_ele} Diagrammatic elements of Eulerian perturbation 
theory. {\it left}: perturbative expansion (Eq. \ref{eqn:psid}); 
{\it right}: initial power spectrum $P_0(k)$. 
}
\end{figure}

The power spectrum $P_m(k,\eta)$ of the matter density perturbation is defined as
\begin{eqnarray}
 P_m(k,\eta) \delta_D(\vk+\vk^{\pr}) =  
 \left \langle \Xi_m(\vk,\eta) ~ \Xi_m(\vk^{\pr},\eta) \right \rangle
\end{eqnarray}
For Gaussian initial conditions, all the statistical
information is encoded in the initial power spectrum
\begin{eqnarray}
\label{eqn:init_P}
 \langle \phi_a(\vk) \phi_b(\vk^{\pr}) \rangle = \delta_D(\vk+\vk^{\pr})
  P_{ab}(k),
\end{eqnarray}
where $P_{ab}(k)=u_a u_b P_0(k)$, with 
$P_0 (k) = \langle \delta_0(\vk)\delta_0(\vkpr) \rangle$.
Diagrammatically, as shown in the right-hand diagram of Fig. (\ref{fig:EPT_mat_ele}), the ensemble average is 
obtained by gluing two open circles together to form the symbol $\otimes$, 
which represents the initial power spectrum $P_{ab}(k)$.

\begin{figure*}[!htp]
\begin{center}
\includegraphics[width=0.75\textwidth]{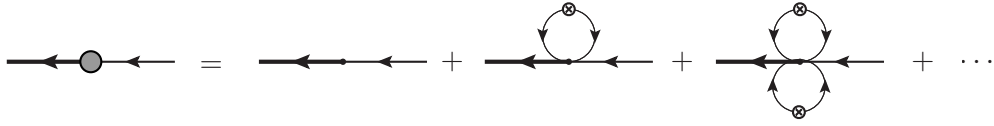}
\end{center}
\caption{\label{fig:P_psi} The nonlinear propagator $\Gamma^{(1)}_m(k,\eta)$ 
 has an infinite number of loop contributions.
}
\end{figure*}

As the density fluctuations evolve into the non-linear regime at later times, 
the validity of standard perturbation theory breaks down,  loop 
contributions become ill-behaved and the convergence of perturbation series 
gets out of control. As one of several different approaches beyond SPT been proposed recently, we will briefly review the renormalized perturbation theory (RPT)
introduced by \cite{CS06a,CS06b,BCS08}, 
and then the Lagrangian resummation theory \citep{M08a} in the next subsection.

The crucial step of renormalized perturbation theory is to define
the generalized growth factor, known as the propagator $\Gamma^{(1)}_{m ~ab}(k)$
\begin{eqnarray}
\label{eqn:Gab}
\Gamma^{(1)}_{m~ ab}(k,\eta) \delta_D(\vk-\vk^{\pr}) 
   \equiv \left \langle \frac{\delta \Xi_a(\vk,\eta)}
   {\delta \phi_b(\vk^{\pr})} \right \rangle,
\end{eqnarray}
which effectively describes the time evolution of individual Fourier modes when 
non-linear mode-coupling is included. Here $\delta$ denotes the functional
derivative. 
The propagator measures the dependence of a non-linearly evolved Fourier mode 
$\Xi_a(\vk, \eta)$ on its initial state $\phi_b(\vk)$ on average.
Intuitively, one should expect $G_{ab}$ to decay to zero at small scales since 
non-linear mode-coupling has erased all the information from the initial state
at that $\vk$.
Indeed, with the help of the Feynman diagrams introduced in \cite{CS06a}, the 
dominant contribution can be summed up explicitly in the large-k limit, and giving 
the exponential decay $G_{ab}(k) \approx \exp(-k^2 D^2 \sigma_{\Psi}^2 /2)$. 
In this paper, the non-linear 
propagator is diagrammatically presented as a grey circle with an incoming branch.
It is a summation of infinite number of loop contributions, as illustrated in 
Fig. (\ref{fig:P_psi}). 

Given above, \cite{CS06a} was able to rewrite the nonlinear matter density power 
spectrum as a sum of two contributions
\begin{eqnarray}
P_m(k,\eta) = G^2(k, \eta) P_0(k) + P_{MC}(k,\eta),
\end{eqnarray}
where $G$ is the density propagator $G(k,\eta)=G_{1b}u_b$, and $P_{MC}(k,\eta)$ is 
the mode-coupling term. 
Therefore, the non-linear power spectrum at any $\vk$ is composed of two parts.
One is proportional to the initial power spectrum at the same $\vk$; the other 
comes from mode-coupling of other $\vk^{\pr}$. As $G(k)$ decays at small scales,
more and more power comes from the mode-coupling contribution.
\cite{BCS08} showed that these complicated mode-coupling contributions
can be expressed as a summation of multi-point propagators, defined as
\begin{eqnarray}
\label{eqn:gamma_psi}
\Gamma^{(n)}_{m ~a b_1 \cdots b_n}  (\vk_1,\cdots,\vk_n; \eta) 
\delta_D(\vk-\vk_{1\cdots n})  \nonumber \\
 \qquad = \quad \frac{1}{n!} ~ \left \langle 
\frac{\delta^n \vxi_a(\vk, \eta)}{\delta 
\phi_{b_1}(\vk_1) \cdots \delta \phi_{b_n}(\vk_n)  } \right \rangle,
\end{eqnarray}
which is nothing but a generalization of two-point propagator $G_{ab}$.
Similarly, the dominant part of multi-point propagators 
can also be summed and decay into the nonlinear regime at the same rate as 
two-point propagator.
\cite{BCS08} showed that a simple approximation which generalizes the
$k$-dependence of two-point propagators agrees with the data with acceptable 
accuracy,
\begin{eqnarray}
 \label{eqn:gammpsi_apprx}
 \Gamma^{(n)}_m (\vk_1,\cdots,\vk_n ;\eta) & =& 
 \Gamma^{(n,~{\rm tree})}_m(\vk_1,\cdots,\vk_n;\eta)\nonumber \\
&& \times ~~\frac{\Gamma^{(1)}_m(|\vk_{1\cdots n}|) }{\Gamma^{(1,~{\rm tree})}_m
   (|\vk_{1\cdots n}|) }.
\end{eqnarray} 
For gaussian initial condition, the nonlinear power spectrum can then be expressed 
with the help of Eq.(\ref{eqn:gamma_psi}), known as the $\Gamma-$expansion
\begin{eqnarray}
\label{eqn:P_psi}
P_{m~ ab}(k, \eta) &=& \sum_{n \ge 1} n! \int d^3\vq_{1 \cdots n} ~ 
\delta_D(\vk - \vq_{1 \cdots n})  \nonumber \\
&& ~~\Gamma_{m ~a }^{(n)}(\vq_1,  \cdots, \vq_n; \eta) 
~ \Gamma_{m ~b }^{(n)}(\vq_1,\cdots, \vq_n; \eta) \nonumber \\
&&  ~~ P_0(q_1) \cdots P_0(q_n).
\end{eqnarray}

Note that Eq. (\ref{eqn:P_psi}) describes nothing but an alternative way of
taking ensemble averages, or diagrammatic speaking, gluing initial states.
Instead of gluing two density fields order by order, 
one can first construct the objects by gluing initial states of individual
density fields with $n$ incoming branches, known as an ($n+1$)-point propagator, 
and the final non-linear power spectrum is then obtained by gluing two propagators 
together. 
Therefore, this $\Gamma-$expansion is quite general and applicable to the ensemble 
average of any two arbitrary statistical fields $\langle x_1 x_2 \rangle$, 
which are functionals of some other field $x_0$
\begin{eqnarray}
    \label{eqn:xi_def}
    x_i[x_0(\vk_n), \lambda] &=& \sum_n \int d^3\vk_{1\cdots n} ~ X^{(n)}_i (\vk_1,\cdots,\vk_n; \lambda) \nonumber \\  
    && \qquad\times~  [x_0(\vk_1) \cdots x_0(\vk_n)], 
\end{eqnarray}
as long as the building blocks are well estimated. 
Here $\lambda$ symbolize some parameters, and $X^{(n)}$ are symmetric on their arguments.
In the case that $x_0$ is Gaussian, the Wick's theorem ensures that the ensemble 
average of an odd number of fields vanishes
$ \langle x_0(1)\cdots x_0(2n+1)\rangle =0$ and the average of an even number of fields
 $\langle x_0(1)\cdots x_0(2n) \rangle $ can be 
 decomposed as summation of two-point correlations
\begin{eqnarray}
    \label{eqn:wick}
    \langle x_0(1)\cdots x_0(2n) \rangle = 
    \sum_{\rm all~pairs} \prod_{ij}\langle x_0(i) x_{0}(j) \rangle.
\end{eqnarray}
Following the idea of \cite{BCS08}, there are three types of two-point correlations
$\langle x_0(i) x_0(j)\rangle$ involved in calculating $\langle x_1(\lambda) x_2(\lambda^{\pr}) \rangle$: 
pairs within $x_1$ or $x_2$ themselves and pairs connecting them.
Therefore, one is always able to regroup Eq.(\ref{eqn:wick}) in the following way
\begin{eqnarray}
    \label{eqn:wick_regroup}
    \langle x_0(1)\cdots x_0(2n) \rangle = \sum_{\rm in ~between}
    \biggl[\sum_{x_1~ {\rm internal}} \prod_{lm} 
     \langle x_0(l)x_0(m)  \rangle  \biggr]  \nonumber \\ 
    \biggl [\sum_{x_2~{\rm internal}}      \prod_{l^{\pr}m^{\pr}} \langle x_0(l^{\pr})x_0(m^{\pr})  \rangle  \biggr]  
    \biggl[ \prod_{ij~{\rm in~ between}} \langle x_0(i)x_0(j)  \rangle  \biggr].
    \nonumber \\
\end{eqnarray}
Combining above equations (\ref{eqn:xi_def}, \ref{eqn:wick} and \ref{eqn:wick_regroup}) 
together, one is able to reformulate $\langle  x_1(\lambda) x_2(\lambda^{\pr}) \rangle$
as a sum of two-point cross correlations of $x_0$ belonging to $x_1$ and $x_2$ respectively
with some pre-summed nonlinear quantity involving internal pairs 
$\sum_{\rm internal} \prod_{lm}\langle x_0(l)x_0(m) \rangle$, which  can be formally 
defined as the average of functional derivative 
\begin{eqnarray}
\Gamma_i^{(n)} =\frac{1}{n!} \left \langle \frac{  \delta^n x_i}{\delta x_0(1)
  \cdots \delta x_0(n) }\right\rangle.
\end{eqnarray}
Therefore, as long as $\Gamma_i^{(n)}$ is well estimated, one is always able to 
carry out the $\Gamma-$expansion in such situation.
In the following, we will demonstrate that
with the help of the density propagator of cold dark matter (Eq. \ref{eqn:gamma_psi}), 
many similar quantities of this kind can be calculated. 
And we will extensively utilize such expansion in the following of the paper.

\begin{figure}[!htp]
\begin{center}
\includegraphics[width=0.3\textwidth]{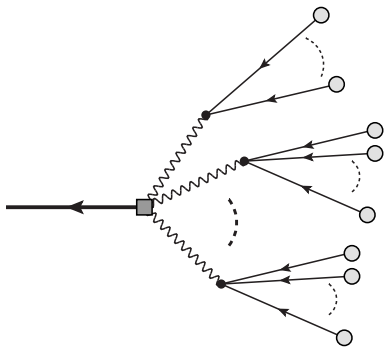}
\end{center}
\caption{\label{fig:LPT_mat_ele} Diagrammatic representation of nonlinear matter 
density perturbation in Lagrangian perturbation theory, similar to the one 
introduced in \cite{M08a}. The wavy line denotes the displacement field $\vpsi$, 
and the thin solid lines together with grey open circles stand for the linear 
density perturbations $\delta_0$.  
}
\end{figure}

\subsection{Lagrangian Perturbation Theory}
As discussed in the beginning of this section, the dynamics of the system 
in Lagrangian picture is fully characterized by the displacement field 
$\vpsi(\vq,\eta)$. 
Combining the equation of motion (Eq. \ref{eqn:LPT_dynamics}) and the Poisson equation,
the displacement field can be 
solved perturbatively. In Fourier space, the $n-$th order perturbation
$\vpsi^{(n)}$ can be generally expressed as 
\begin{eqnarray}
    \label{eqn:LPT_expansion}
    \vpsi^{(n)}(\vk) =  \frac{i D^n}{n!}\int d^3\vp_{1\cdots n} ~\vL^{(n)}(\vp_{1\cdots n}) 
    \delta_0(\vp_1) \cdots \delta_0(\vp_n),\nonumber \\
\end{eqnarray}
where $\vL^{(n)}$ is the perturbation kernel.

Due to the matter conservation, the Eulerian density contrast $\delta_m(\vx)$ 
 can be expressed as
\begin{eqnarray}
    \label{eqn:LPT_deltam}
    1+\delta_m(\vx) = \int d^3\vq \left [ 1+\delta_0(\vq) \right ]
     \delta_D(\vx-\vq-\vpsi(\vq)).
\end{eqnarray}
For the matter distribution, one usually assumes that the initial field is 
sufficiently uniform 
$\delta_0 \approx 0$, so that above equation (Eq.\ \ref{eqn:LPT_deltam}) can be 
further simplified. In Fourier space,
\begin{eqnarray}
    \label{eqn:LPT_dentam_Fourier}
    \delta_m (\vk) &=& \int d^3\vq~ e^{-i\vk\cdot \vq} ~\left [e^{-i\vk\cdot \vpsi(\vq)}-1
    \right]\nonumber\\
    &=&  \sum_{n=1}^{\infty} \frac{(-i)^n}{n!} \int d^3\vp_{1\cdots n} \left[\vk 
    \cdot \vpsi(\vp_1) \right] \cdots  \left[\vk \cdot \vpsi(\vp_n) \right] 
      \nonumber \\ && \qquad\qquad \qquad  \times~~ \delta_D(\vk-\vp_{1\cdots n}).
\end{eqnarray}
Similarly, the above equation can also be represented by diagrams as shown in 
Fig.\ (\ref{fig:LPT_mat_ele}). Here, wavy lines denote the displacement field $\vpsi$,
and the thin solid line together with grey open circle stands for the linear 
density perturbation, i.e. Eq.(\ref{eqn:LPT_expansion}).

\begin{figure}[!htp]
\begin{center}
\includegraphics[width=0.45\textwidth]{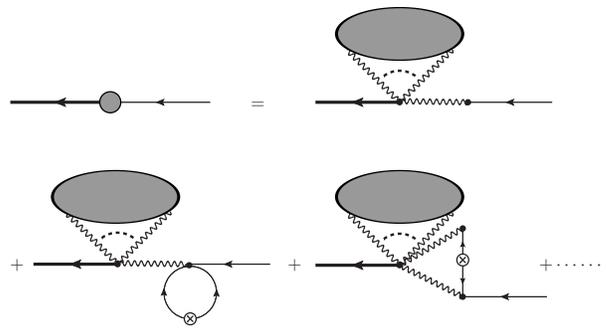}
\end{center}
\caption{\label{fig:LPT_G1} Two-point propagator of matter density $\Gamma_m^{(1)}(\vk)$
 in the context of Lagrangian perturbation theory up to one-loop order. 
 The grey ellipse shown in each contribution symbolizes the quantity of 
 Eq (\ref{eqn:Pi_def}). 
 Note here we have omitted the  summation sign over the displacement 
 field $\vpsi$ presented in \cite{M11}.
}
\end{figure}

Starting from Eq.(\ref{eqn:LPT_dentam_Fourier}), \cite{M08a} performed a resummation
of the power spectrum of the density contrast, known as Lagrangian Resummed 
Perturbation Theory.
Although not been explicitly discussed in \cite{M08a}, it is also possible to define 
the same propagator within this framework. 
First of all, as pointed out in \cite{M11} for biased tracers\footnote{Setting $b_1=1$ 
and $b_n=0$ otherwise in Eq. (71) of \cite{M11}, one recovers the following equation.}, 
the following quantity can be resummed
\begin{eqnarray}
    \label{eqn:Pi_def}
 \sum_{n=0}^{\infty} \frac{(-i)^n}{n!} \int d^3\vp_{1\cdots n} ~ [k_{i_1}\cdots k_{i_n}]
   \left \langle  \Psi_{i_1}(\vp_1) \cdots \Psi_{i_n}(\vp_n)
    \right  \rangle \nonumber \\
    = \sum_{n=0}^{\infty}  \frac{(-i)^n}{n!} \left \langle \left[  \int d^3 \vp~
       \vk \cdot \vpsi (\vp)   \right]^n \right \rangle 
    = \langle e^{-i\vk\cdot \vpsi(0)} \rangle, \qquad
\end{eqnarray}
where $\vpsi(0) = \int d^3\vp ~\vpsi(p)$ is the displacement field at the origin.
It is represented as a grey ellipse in Fig. (\ref{fig:LPT_G1}), and can be further
simplified as \cite{M08a,M11}
\begin{eqnarray}
    \label{eqn:LPT_Pi0}
 \left\langle  e^{-i\vk\cdot \vpsi(0)}  \right\rangle &=& \Pi^{(0)}(\vk) 
    = \exp\left [ \left \langle e^{-i\vk\cdot \vpsi(0)} \right 
    \rangle_c \right] \nonumber \\
   & = &  \exp\left[ \sum_{n=1}^{\infty} \frac{(-1)^n \langle |\vpsi(0)|^{2n} \rangle_c}  
 {(2n+1)(2n)!} k^{2n}   \right ]  \nonumber \\
   & \approx & \exp \left [ -\frac{k^2 D^2}{2}   \sigma_{\Psi}^2 \right] 
\end{eqnarray}
with the variance of displacement field along one direction 
$D^2 \sigma^2_{\Psi} = \langle|\vpsi(0)|^2 \rangle/3 = \int d^3\vq D^2 P_0(q)/(3q^2)$.
Here, we have used the cumulant theorem
\begin{eqnarray}
    \label{eqn:cumulant}
    \left \langle e^{ \left[\sum_{n\ge 0} j_n x_n \right] } \right \rangle =
    \exp \left[\left \langle e^{ \left[\sum_{n\ge 0} j_n x_n \right] } -1
     \right \rangle_c \right],
\end{eqnarray}
where $\langle \cdots \rangle_c$ denotes the connected part of the ensemble average.
As shown in \cite{M11}, it leads us to a partial resummation of the propagator as
 represented in Fig. (\ref{fig:LPT_G1}) up to the one-loop order. 
 Again, this is just a simplified
version of Fig. (17) in \cite{M11} for the Lagrangian bias model.
The grey ellipse represents all possible graphs attached with various number of 
wavy lines, i.e. Eq. (\ref{eqn:Pi_def}). 
Compared to the diagram introduced in \cite{M11}, we have omitted the 
summation sign over external wavy lines. 
Mathematically, such diagrammatical series is equivalent to 
\begin{eqnarray}
    \label{eqn:LPT_G1}
  \Gamma_m^{(1)} (\vk) 
    &=& \left \langle e^{-i\vk\cdot \vpsi(0)} \int d^3\vp \left[ -i \vk \cdot 
   \frac{\delta}{\delta \phi(\vkpr)} \vpsi(\vp)  \right] \right \rangle
     \nonumber\\
    &=& \Pi^{(0)}(\vk) \left \{ \left \langle \frac{\delta}{\delta \phi(\vk^{\pr})}
    \exp \left [ -i \vk \cdot \vpsi(0)\right ]  \right \rangle_c \right \} \nonumber \\
\end{eqnarray}
Note that we have used an alternative version of the cumulant theorem in Eq. 
(\ref{eqn:cumulant}).
 \begin{eqnarray}
     \left \langle e^{ j_1 x_1 } x_2 \right \rangle = 
     \exp \left[\left \langle e^{ j_1 x_1} -1
      \right \rangle_c \right] \left \{ \left \langle e^{ j_1 x_1 } x_2 
      \right \rangle_c  \right \}.
 \end{eqnarray}
which is obtained by taking the derivative of Eq. (\ref{eqn:cumulant}) 
respect to $j_2$ and then setting it to zero.
We further calculate these contributions explicitly in the Appendix A.  
The result is not the same as the form originally proposed by \cite{CS06b}.
However, the matching prescription between high-k and one-loop calculation in 
\cite{CS06b} seems more or less ad hoc. 
Therefore, an alternative way of constructing the propagators was proposed by
\cite{B11}, to which our calculation do agree.
See \cite{B11} for the discussion of different versions of calculating multipoint
propagators.

\begin{figure}[!htp]
\begin{center}
\includegraphics[width=0.35\textwidth]{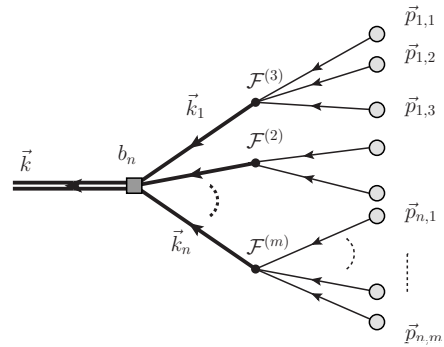}
\end{center}
\caption{\label{fig:phid} Diagrammatic representation of 
$\delta_E(\vk, \eta)$ in Eulerian bias model.
}
\end{figure}

\section{The Non-linear Bias Model}

\subsection{Eulerian Bias Model}
In Eulerian bias model, the galaxy number density perturbation at time $\eta$
is considered as some general functional $F[\delta_m]$ of the underlying matter density
distribution characterized by its Taylor coefficient $b_n$
\begin{eqnarray}
    \delta_E(\vx,\eta) = F\left[  \delta_m(\vx,\eta) \right] = \sum_{n} \frac{b_n}{n!} 
   \left[\delta_m(\vx, \eta) \right]^n,
\end{eqnarray}
where the nonlinear mass density field $\delta_m(\vx)$ is solved by perturbation theory
mentioned in the last section. As discussed by \cite{M11}, this functional is generally
non-local. Therefore in Fourier space, bias coefficient $b_n$ depends on wavenumbers
\begin{eqnarray}
\label{eqn:Ebias}
\delta_E(\vk,\eta) &=& \sum_{n=1}^{\infty} \frac{1}{n!} \int d^3\vk_{1 \cdots n}
~ \delta_D(\vk-\vk_{1\cdots n}) b_n(\vk_1, \cdots, \vk_n)  \nonumber\\
&&\qquad \times~ \left [\delta_m (\vk_1, \eta)
  \cdots   \delta_m(\vk_n, \eta) \right].
\end{eqnarray}

In the local bias model, $b_n$ then becomes constant times a smoothing window
function $W(k, R)$ in Fourier space.
In this model, the linear bias relation at large scale $P(k)\approx b^2 P_m(k)$ 
is not only attributed to the first term in Eq. (\ref{eqn:Ebias}), but also 
from the higher orders $b_n$.
As first pointed out by \cite{SW98}, the effective large scale bias is determined 
by infinite number of higher-point local variances of the field.
Therefore, similar to the situation in the standard perturbation theory of 
matter density field, at later time when mass density variance grows larger, 
a naive perturbative expansion fails. 
In this case, such a breakdown affects both the large scale amplitude and the 
high-k regime of the power spectrum.

Given the definition of $\delta_E(\vk)$ in Eq.(\ref{eqn:Ebias}) and the 
diagrammatic representation of the density perturbation in Fig. (\ref{fig:EPT_mat_ele}), 
it is also possible to draw diagrams representing $\delta_E(\vk)$. 
As depicted in Fig.(\ref{fig:phid}), each diagram contains two 
levels of nonlinear interactions.  The thick solid $n$-branch tree represents the 
nonlinear convolution, and their interaction vertex, shown as a solid square, carries 
the bias coefficient $b_n/n!$, where $n$ is
the number of branches.
Each branch represents one nonlinearly evolved density contrast
$\delta_m(\vk_n, \eta)$ at time $\eta$, which is followed by another $m$ thin
branches, to express its gravitational nonlinearity in terms of initial fields and 
perturbative kernels $\mF^{(m)}$.
At the end of each branch, small open circles represent the initial conditions
$\delta_0$.  When expanding $\delta_E(\vk, \eta)$ to different orders, 
one should also include the information about the number of topologically 
equivalent diagrams, which essentially comes from the
multinomial coefficient of $(\delta_m)^n$. 
The final field $\delta_E(\vk, \eta)$ is then a summation 
of all possible diagrams.

\subsection{Lagrangian Bias Model}
\begin{figure}[!htp]
\begin{center}
\includegraphics[width=0.5\textwidth]{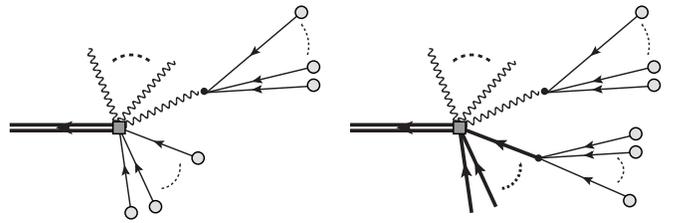}
\end{center}
\caption{\label{fig:LPT_bg_ele} Comparison between Lagrangian bias model 
({\it left}) discussed in \cite{M08a} and our galaxy formation model 
({\it right}).  The thin solid denotes the linear density perturbation $\delta_0$,
and thick line represents nonlinear overdensity $\delta_m(\eta)$. Note that the 
right diagram has not revealed very clearly the process of continuous galaxy formation 
as shown in Eq. (\ref{eqn:deltag_c}).  
}
\end{figure}

To incorporate the halo bias with nonlinear perturbation theory, \cite{M08b}
proposed an alternative bias model involving the initial Lagrangian density 
fluctuation $\delta_0(\vq)$ instead of final Eulerian density $\delta_m(\vx, \eta)$.
In this Lagrangian bias model, the non-linear objects, typically halos, were 
formed at some initial time with the number density is
a functional $F[\delta_0(\vq)]$ of underlying linear density distribution 
$\delta_0$. 
Then the Eulerian observable $\delta_L(\vx, \eta)$ at later time is described
by the continuity relation
\begin{eqnarray}
    \label{eqn:Lbiasing}
    1+\delta_L(\vx) &=& \int d^3\vq \left [ 1+\delta_L(\vq)  \right ]
    \delta_D[\vx-\vq-\vpsi_L(\vq)] \nonumber \\
    &=& \int  d^3\vq ~ F[\delta_0(\vq)] ~
    \delta_D[\vx-\vq-\vpsi_L(\vq)], \nonumber \\
\end{eqnarray}
where $\vpsi_L(\vq)$ is the displacement field of such tracer. In Fourier space,
\begin{eqnarray}
    \label{eqn:Lbias_k}
  \delta_L(\vk,\eta) = \int d^3\vq~e^{-i\vk\cdot \vq} \left [ F[\delta_0(\vq)] 
    ~ e^{-i\vk\cdot \vpsi(\vq,\eta) } -1 \right].
\end{eqnarray}
In the limits that nonlinear objects trace exactly the underlying dark matter, 
i.e. $\vpsi_L = \vpsi_m$,
one can further proceed to calculate the power spectrum $P(\vk)$ using Lagrangian
perturbation theory of the matter field.
In Fig. (\ref{fig:LPT_bg_ele}), we reproduced the diagram of this model
introduced by \cite{M11} on the left, where the wavy line represents the displacement field
$\vpsi_L$ and the $n$ branches of thin solid lines with open circle symbolize the
$n$th order of $F[\delta_0]$.

The great advantage of this approach is that it naturally takes account of the non-locality
induced by gravitational evolution. As shown by \cite{M11}, after perturbatively expanding
both sides of Eq.(\ref{eqn:Lbias_k}) and comparing terms order by order, 
one obtains the relationship between Eulerian and Lagrangian bias parameters. 
To the first two orders \citep{M11}, 
\begin{eqnarray}
    b^E_1 (\vk) &=& 1+ b^L_1 (\vk), \nonumber \\
    b^E_2 (\vk_1,\vk_2) &=& b^L_2(\vk_1,\vk_2) - F_2(\vk_1,\vk_2) b^L_1(\vk)
    + \left [ \vk\cdot L_1(\vk_1) \right]    \nonumber \\ 
   &&  \times ~ b^L_1(\vk_2)   + \left[ \vk\cdot L_1(\vk_2) \right] b^L_1(\vk_1)    . 
\end{eqnarray}
Therefore, even if the Lagrangian bias $b^L_n$ is local initially, non-linear and
non-local gravitational evolution $\mF^{(n)}(\vp_1,\cdots,\vp_n)$ would render the 
Eulerian bias $b^E_2$ non-local.

\subsection{Continuous Galaxies Formation Model}
As briefly mentioned in the introduction, we will try to generalize the Eulerian bias
model in two different ways. As the resummation technique will be discussed in next two 
sections, in the following of this section,  we will try to construct a  
model by combining Eulerian and Lagrangian pictures together.

First, we still assume that the complicated nonlinear process leading to
the formation of galaxies or other type of observable tracers can be described
by a general nonlinear functional $F[\delta_m(\eta)]$ at the moment of galaxy creation.
In general this can be nonlocal, however, we will assume $F[\delta_m]$
to be local for simplicity. 
As what will be seen in the following, the locality assumption here is not as
problematic as in the local Eulerian bias model since it only relates to the physical
process during the moment of galaxy formation and could be easier to generalize. 
Furthermore, such functional can also explicitly depends on the formation
time, i.e. we have $F[\delta_m(\eta),\eta]$ and $b_n(\vk_{1\cdots, n},\eta)$.

Assuming at time $\eta$,  there are $\Delta \rho_g (\vq, \eta)$ new galaxies were 
created during $\Delta \eta$ at a Lagrangian-like position $\vq$, and then evolved to 
Eulerian position $\vx$ where have been observed by a survey at time $\eta_0$. 
Analogous to the Lagrangian bias model (Eq. \ref{eqn:Lbiasing}), this contributes 
to the Eulerian galaxies a number density $\rho_g(\vx,\eta_0)$
\begin{eqnarray}
    \label{eqn:cont_deltarho}
    \int d^3\vq ~\Delta \rho_g (\eta,\vq)~ 
    \delta_D[\vx-\vq-\vpsi_g(\vq, \eta, \eta_0)],
\end{eqnarray}
where $\vpsi_g(\vq,\eta,\eta_0)$ is the displacement field of galaxy from $\eta$ 
to $\eta_0$. Note that $\Delta \rho$ here only comprises galaxies that are finally
selected in the sample. 
Therefore, the galaxy number density at observed time $\eta_0$ and comoving position $\vx$ 
is a sum of all contributions that formed before $\eta_0$
and then moved to $\vx$ with a displacement $\vpsi_g(\vq,\eta,\eta_0)$,
\begin{eqnarray}
    \label{eqn:cont_rhog}
    \rho_g(\vx,\eta_0) &=& \int_{\eta_{min}}^{\eta_0} 
    d\eta \int d^3\vq ~ \bar{\rho}^{\pr}_g(\eta) F[\delta_m(\vq,\eta),\eta] ~
    \nonumber\\ && ~\times ~
    \delta_D[\vx-\vq-\vpsi_g(\vq, \eta, \eta_0)].
\end{eqnarray}
Here we have written the number density 
$\Delta \rho_g (\vq,\eta) = \overline{\Delta  \rho}_g 
(\eta) F[\delta_m(\eta),\eta]$, where $\delta_m(\vq,\eta)$ is the nonlinear evolved
matter density field at $\eta$. In principle, the displacement field of 
galaxies $\vpsi_g$ should follow the one of dark matter $\vpsi_m$, i.e.
\begin{eqnarray}
    \label{eqn:xig}
    \vpsi_g(\vq; \eta, \eta_0) = b_{\Psi} \left [\vpsi_m(\vq_{in},\eta_0) - 
    \vpsi_m(\vq_{in},\eta) \right],
\end{eqnarray}
where $\vq_{in}$ is some initial Lagrangian position of the matter field when density 
perturbation is still linear. In the following of this paper, we will assume
$b_{\Psi} = 1$.
In the simplest case where all galaxies formed at $\eta_f$, i.e. 
$\bar{\rho}^{\pr}_g (\eta) =\bar{\rho}_g \delta_D(\eta- \eta_f)$, a similar form 
of Eq.(\ref{eqn:Lbiasing}) is recovered.
Taking the average of both sides of Eq.(\ref{eqn:cont_rhog}), one obtains the 
average number density $\bar{\rho}_g(\eta_0)$ as a integration of the average
density over the whole galaxy formation history
\begin{eqnarray}
    \bar{\rho}_g(\eta_0) = \int_{\eta_{min}}^{\eta_0} d\eta ~\bar{\rho}^{\pr}_g(\eta)
     \langle F[\delta_m,\eta]     \rangle   
     = \int_{\eta_{min}}^{\eta_0} d\eta~  \bar{\rho}^{\pr}_g(\eta)
\end{eqnarray}
Therefore, the perturbation of galaxy number density can then be expressed as,
\begin{eqnarray}
    \label{eqn:deltag_c}
     \delta_g(\vk,\eta_0) &=& \int_{\eta_{min}}^{\eta_0} d\eta \int d^3\vq~
    f(\eta)~ e^{-i\vk\cdot \vq} F[\delta_m(\vq,\eta),\eta] \nonumber\\ && ~\times ~
    e^{-i\vk\cdot \vpsi_g(\vq,\eta,\eta_0)} - \delta_D(\vk).
\end{eqnarray}
where $f(\eta) = \bar{\rho}^{\pr}_g (\eta) /\bar{\rho}_g(\eta_0) $ denotes the
 average galaxies formation rate, with the normalization relation
  $\int d\eta ~f(\eta) = 1$.
To first order, we have
\begin{eqnarray}
    \delta_g(\eta_0) &=&  \int_{\eta_{min}}^{\eta_0} d\eta ~ f(\eta)
    \left[ b_1 D(\eta) + b_{\Psi} \left[D(\eta_0)-D(\eta)\right]\right]\delta_0 \nonumber \\
    &=&\left[ b_{\Psi} D(\eta_0) + \int_{\eta_{min}}^{\eta_0}d\eta ~f(\eta) D(\eta) (b_1-
    b_{\Psi}) \right] \delta_0,\nonumber \\
\end{eqnarray}
Therefore, the linear bias of galaxies evolves as 
\begin{eqnarray}
    \label{eqn:linear_debias}
    b^g_1(\eta_0) = b_{\Psi} + \int_{\eta_{min}}^{\eta_0}d\eta ~f(\eta) \frac{D(\eta)}
    {D(\eta_0)} [b_1- b_{\Psi}].
\end{eqnarray}
In the case that galaxies exactly trace cold dark matter, i.e. $b_{\Psi}=1$, we recover
the result calculated by \cite{CSS12} if only the growing mode is considered. 
As first discussed by \cite{TP98}, for the model galaxies burst at a single time 
$\eta_f$, $f(\eta)=\delta_D(\eta-\eta_f)$, 
\begin{eqnarray}
    b^g_1(\eta_0) = 1 + \frac{D(\eta_f)}{D(\eta_0)} [ b_1 - 1].
\end{eqnarray}
So the gravitational growth will always debias the clustering of galaxies towards 
$b^g_1 = 1$.

Because of this Lagrangian-like approach adopted here, this framework automatically
incorporates the gravitational non-locality induced after the formation of the 
galaxies.
Furthermore, as will be seen in Section V., Eq. (\ref{eqn:deltag_c}) also allows
the application of the resummation technique so that the perturbative convergence
of such model is guaranteed.

\section{Resummed Perturbation Theory of Eulerian Bias Model}

\begin{figure}[!htp]
\begin{center}
\includegraphics[width=0.5\textwidth]{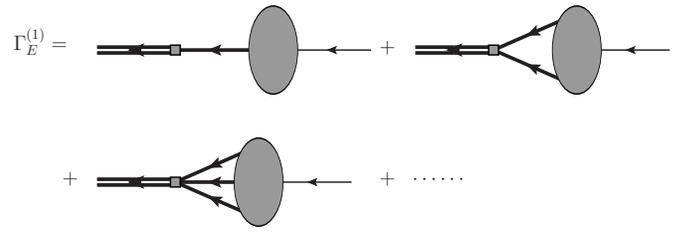}
\end{center}
\caption{\label{fig:GA1} Diagrammatical demonstration of the two-point nonlinear 
propagator $\Gamma^{(1)}_E(k)$ for Eulerian bias mode. The grey ellipse includes
all possible ensemble averages among initial density perturbation $\delta_0$. 
}
\end{figure}

In this section, we will show that the Eulerian bias model can be partially
resummed with the help of exponential decay of multipoint matter propagators
 $\Gamma_m^{(n)}$. Starting from the definition of this model (Eq. \ref{eqn:Ebias}),
one can define the ($n+1$)-point nonlinear propagator 
$\Gamma^{(n)}_E$ of the biased-tracer overdensity $\delta_E$,
\begin{eqnarray}
\Gamma^{(n)}_{E,~b_1 \cdots b_n} (\vk_1,\cdots,\vk_n; \eta) 
\delta_D(\vk-\vk_{1\cdots n})  \qquad \nonumber \\
 = \quad \frac{1}{n!} ~\left \langle 
\frac{\delta^n \Xi_E(\vk)}{\delta \phi_{b_1}(\vk_1) \cdots \delta 
\phi_{b_n}(\vk_n)  } \right \rangle.
\end{eqnarray}
Before substituting $\delta_E$ (Eq. \ref{eqn:Ebias}) as well as the perturbative 
solution of $\delta_m$ (Eq. \ref{eqn:gamma_psi}, \ref{eqn:P_psi}) into above 
definition, one first notices that, $\Gamma^{(n)}_E$ can be expanded in terms of
the number of nonlinear density fields $\delta_m$ involved, just as 
depicted in the schematic diagram of Fig.(\ref{fig:GA1}) for the two-point 
propagator. Mathematically, it means $\Gamma^{(n)}_E= \sum_m \Gamma^{(n,~m)}_E$, 
where
\begin{eqnarray}
    \label{eqn:Gn_m_def}
    \Gamma^{(n,~m)}_{E,~b_1 \cdots b_n}(\vk_1,\cdots,\vk_n;\eta) 
    \delta_D(\vk-\vk_{1\cdots n}) \qquad \qquad \qquad \nonumber \\
    = \frac{1}{n!m!}\int d^3\vq_{1\cdots m}
    ~\delta_D(\vk-\vq_{1\cdots m}) b_m(\vq_1,\cdots, \vq_m)\nonumber\\
    ~  \left \langle \frac{\delta^n}{\delta \phi_{b_1}(\vk_1)\cdots 
  \delta \phi_{b_n}(\vk_n)} 
    \left [ \Xi_m(\vq_1) \cdots \Xi_m(\vq_m) \right ] \right \rangle.
\end{eqnarray}

For the two-point propagator $\Gamma_E^{(1)}(\vk)$,  let us write down all 
contributions formally.
First of all, it is clear that the one-$\delta_m$ term, i.e.\ the first 
diagram in Fig.(\ref{fig:GA1}), equals the two-point propagator of the matter
field itself $\Gamma_m^{(1)}(\vk)$ multiplied with the linear bias parameter
 $b_1(k)$. For the two-$\delta_m$ contribution, i.e.\ the second diagram, one reads
 from Eq.\ (\ref{eqn:Gn_m_def})
\begin{eqnarray}
    \label{eqn:G1_2_def}
    \Gamma^{(1,~2)}_{g, ~a}(\vk; \eta)&=& 
      2  \int d^3 \vq_{12} ~\delta_D(\vk-\vq_{12}) 
    ~ \frac{b_2(\vq_1, \vq_2) }{2}~  \nonumber \\
    && ~\times ~ \left \langle
     \frac{\delta \psid(\vq_1,\eta) }{ \delta \phi_a} \psid(\vq_2, \eta) 
     \right \rangle
\end{eqnarray}
As discussed in Section II, the ensemble average at the second line can be 
expanded by a $\Gamma-$like series, which in this case, is the propagator
for matter field $\Gamma^{(n)}_m$. To proceed, we expand Eq. (\ref{eqn:G1_2_def})
explicitly
\begin{widetext}
\begin{eqnarray}
\label{eqn:G1_2}
\Gamma^{(1,~2)}_{E, ~a}(\vk; \eta)
&=&\int d^3\vq_{12} ~\delta_D(\vk-\vq_{12}) ~ b_2(\vq_1, \vq_2)
\sum_{n_1, n_2}  (n_1+1)  \int d^3\vp_{1 \cdots n_1 }
 d^3\vp^{\pr}_{1 \cdots n_2} ~
  \delta_D(\vq_1-\vk-\vp_{1\cdots n_{1}} ) \nonumber \\
&&\times ~ \delta_D(\vq_2-\vp^{\pr}_{1\cdots n_2} ) 
~\mF^{(n_1+1)}_{m, ~a c_1\cdots c_{n_1}}(\vk,\vp_1,\cdots,\vp_{n_1}; \eta) 
\mF^{(n_2)}_{m, ~d_1\cdots d_{n_2}}(\vp^{\pr}_1,\cdots,
\vp^{\pr}_{n_2}; \eta) \nonumber \\
&& \times ~\left \langle \phi_{c_1}(\vp_1)\cdots \phi_{c_{n_1}}
(\vp_{n_1}) 
~ \phi_{d_1}(\vp^{\pr}_1) \cdots \phi_{d_{n_2}}(\vp^{\pr}_{n_2}) \right \rangle .
\end{eqnarray}
\end{widetext}

As we have shown, Wick's theorem ensure the decomposition of joint ensemble 
average at the last line into the combinations of two-point correlations.
Each of these terms can be labeled by three indices $r_1,r_2,t$, 
where $n_1=2r_1+t$, $n_2=2 r_2 +t$, i.e.\ we classify all pairs
into three categories: $r_1$ pairs within the first field 
$\phi_{c_i}(\vp_i), ~(1<i<n_1)$;
$r_2$ pairs within the second field $\phi_{d_j}(\vp^{\pr}_j), ~(1<j<n_2)$;
and $t$ pairs in between. The renormalization is then achieved by realizing 
that a pre-summation of $r_1$ and $r_2$ gives rise to one $(t+2)$-point 
and another $(t+1)$-point propagator, therefore
\begin{eqnarray}
\label{eqn:G1_2_final}
\Gamma^{(1,~2)}_{E, ~a}(\vk; \eta) =
 \sum_n (n+1)! \int d^3\vp_{1 \cdots n} ~ P_0(p_1)\cdots P_0(p_n) 
 \nonumber \\   
 \times ~ b_2(\vk+\vp_{1\cdots n}, -\vp_{1\cdots n})
 ~\Gamma^{(n+1)}_{m,~a}(\vk,~\vp_1,\cdots,\vp_n;\eta) \qquad
 \nonumber \\
\times ~ \Gamma^{(n)}_{m}(-\vp_1,\cdots,-\vp_n;\eta),
\qquad\qquad\qquad \qquad \qquad \qquad 
\end{eqnarray}
where we have introduced the notation
\begin{eqnarray}
\Gamma_{m}^{(n)}(\vp_1,\cdots,\vp_n;\eta) = 
\Gamma_{m, ~c_1 \cdots c_n }^{(n)}(\vp_1,\cdots,\vp_n;\eta) 
 ~ u_{c_1} \cdots u_{c_n},  \nonumber\\
\end{eqnarray}
and used the definition of $\Gamma^{(n)}_{m}$ in
terms of the perturbation kernel. 
The coefficient $n!$ then comes from all possible ways of matching $n$ initial
states with another group of $n$ initial states.
For this contribution, the only difference between Eq.(\ref{eqn:G1_2_final}) and
Eq.(\ref{eqn:P_psi}) is the order of the matter propagator been summed. 
For the RPT description of the matter power spectrum, both propagators share the
same order; otherwise, it would be impossible to match the pair.
For the same reason, the order of one propagator in Eq.(\ref{eqn:G1_2_final}) 
is always greater than the other by one, since we have to select one branch out 
before taking the average.

The formula can be further generalized to arbitrary order $n$,
\begin{eqnarray}
\label{eqn:G1n}
\Gamma^{(1,~n)}_{E,~ a} (\vk;\eta) = 
 \sum_{\{t_{ij}\}}   g^{(1,~n)}_{\{ t_{ij} \}}
 \biggl\{ \int \biggl[ \prod_{ 1\leq i<j\leq n} d^3 \vp^{ij}_{1\cdots t_{ij}} \biggl] 
 \qquad \quad \nonumber \\
~\times~ \frac{b_n(\vk, \vp^{ij})}{n!} 
~\Gamma_{m ~a}^{(t_1 + 1)} (\vk;\vp^{11}_{1\cdots t_{11}},
\cdots,  \vp^{1n}_{1\cdots t_{1n}}; \eta) \qquad\quad \nonumber \\
~ \cdots  \Gamma_{m}^{(t_n)} (\vp^{n1}_{1\cdots t_{n1}},\cdots,
\vp^{nn}_{1\cdots t_{nn}}; \eta ) ~  
 \prod_{ 1\leq i<j\leq n} \left[ P_0(p^{ij}) \right]^{t_{ij}}   \biggr\},\nonumber\\
\end{eqnarray}
where the index $t_{ij}$ denotes the number of connections between the $i$-th 
and the $j$-th density field $\delta_m$ (e.g.\ $\{t_{ij}\}=\{t_{12}, t_{13},
t_{14}, t_{23}, t_{24}, t_{34} \}$ for $n=4$), so the total number of the indices
is then $n(n-1)/2$. 
The coefficient $g^{(1,~n)}_{\{ t_{ij}\}}$ is then given by
\begin{eqnarray}
\label{eqn:gtij}
g^{(1,~n)}_{\{t_{ij}\}} &= & n~(t_1+1) 
~ \biggl[ \prod_{1\leq i\leq n} ~{t_i \choose t_{i1}\cdots t_{in} }
\biggr] \nonumber \\
&& \times \biggl[ \prod_{1\leq i<j\leq n} (t_{ij})! \biggr].
\end{eqnarray}
Here, $t_i = \sum_{m} t_{im}$ is the total number
of connections linked between the $i$-th density field $\delta_m$ and others, with
$t_{ij}=t_{ji}$.
Eq.\ (\ref{eqn:gtij}) expresses the products of multinomial coefficients of 
choosing $\{t_{i1}, \cdots, t_{in}\}$ from $t_i$ for each density field,
times all possible permutations within $t_{ij}$ for pair matching.
Each connection carries a momentum $\vp^{ij}_{l}, ~ ( 1<l<t_{ij} )$, which 
characterizes the $l$-th connection between $i$-th $\delta_m$ and $j$-th $\delta_m$.
The integration is taken over momenta of all possible connections among $n$ 
different density fields. 
Because of the ensemble average Eq.(\ref{eqn:init_P}), we have 
$\vp^{ij}_{t_{ij}} = -\vp^{ji}_{t_{ji}}$.
$b_n(\vk,\vp^{ij})= b_n(\vk+\sum_i\vp^{1i}, \sum_i\vp^{2i},\cdots, \sum_i \vp^{ni})$,
and we also used the shorthand notation $ [P_0(p^{ij})]^{t_{ij}}$ for
$P_0(p^{ij}_1) \cdots  P_0(p^{ij}_{t_{ij}})$.

\begin{figure}[!htp]
\begin{center}
\includegraphics[width=0.49\textwidth]{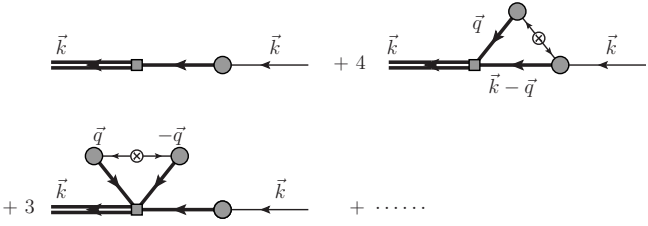}
\end{center}
\caption{\label{fig:GA1_1lp} Nonlinear propagator $\Gamma_E^{(1)}(k)$ 
of Eulerian bias model up to one-loop order. The integer in front of each diagram
indicates the number of all topologically equivalent diagrams.}
\end{figure}

Given above expressions, we are then able to expand $\Gamma^{(1)}_E$ in terms of the
number of initial power spectra entering the calculation. 
For the tree level, it simply reads as
\begin{eqnarray}
 \label{eqn:GA1_tree}
 \Gamma_{E, ~ a}^{(1, ~{\rm tree})} (\vk;\eta) = 
b_1(k) ~ \Gamma_{m, ~a}^{(1)}(\vk;\eta;R) 
\end{eqnarray}
where we have introduced the notation
\begin{eqnarray}
    \Gamma_m^{(n)}(\vp_{1\cdots n};\eta;R) = \Gamma_m^{(n)}(\vp_1 \cdots \vp_n;\eta)
    W(|\vp_{1\cdots n}|R), 
\end{eqnarray}
where $W(kR)$ is some smoothing window function.

At one-loop level, from Eq.(\ref{eqn:G1n}), the contribution would be nonzero 
only when $n\le 3$,
since otherwise there would exist at least one $\Gamma_{m}$ with
zeroth order, which would vanish because 
$\Gamma^{(0)}_{m}= \langle \delta_m \rangle = 0$.
For $n=2$, $t_{12}=1$, the coefficient equals $4$. For $n=3$, 
$t_{23}=1, t_{12}=t_{13}=0$, the coefficient equals $3$. So we have 

\begin{eqnarray}
\label{eqn:GA1_1loop}
 \Gamma_{E,~a}^{(1, ~{\rm 1-loop})} (\vk;\eta) = 
\int d^3\vq~P_0(q) \biggl[ ~ \frac{b_3(\vk,\vq,-\vq)}{2}  \nonumber\\
\times ~  \bigl[\Gamma_{m}^{(1)}(\vq;\eta;R) \bigr]^2 
\Gamma_{m~a}^{(1)}(\vk;\eta;R) 
+2 ~ b_2(\vk-\vq, \vq)  \nonumber \\
 \times ~\Gamma_{m,~a}^{(2)}(\vk, -\vq;\eta;R)
 \Gamma_{m}^{(1)}(\vq;\eta;R)  \biggr]. \qquad\qquad
\end{eqnarray}

\begin{figure}[!htp]
\begin{center}
\includegraphics[width=0.49\textwidth]{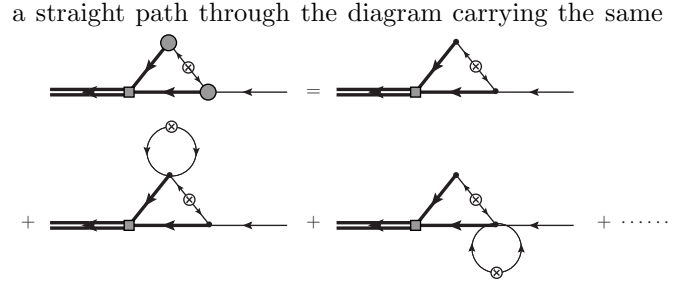}
\end{center}
\caption{\label{fig:GA1_demo} Each diagram contains an infinite number of
loop contributions.}
\end{figure}

To make the Eq.(\ref{eqn:G1n}) easier to understand, one can draw every 
contribution diagrammatically. Starting from the diagram representing 
$\delta_E^{(n)}$ in 
Fig.(\ref{fig:phid}), we change all kernels into $n$-point propagators. 
After selecting one particular branch out, we glue the rest of the initial states 
(open circles) together. 
Every resultant topologically inequivalent diagram represents one or 
several terms in Eq.(\ref{eqn:G1n}).
Since all ordinary kernels have already been substituted by propagators, 
the ensemble average 
(gluing) is only performed among different density fields $\delta_m$.
In Fig.(\ref{fig:GA1_1lp}), we show all diagrams of 
$\Gamma^{(1)}_{E} $ up to one-loop order.  They correspond
one-to-one to Eq.(\ref{eqn:GA1_tree}) and Eq.(\ref{eqn:GA1_1loop}).
We also present all two-loop diagrams and equations of $\Gamma^{(1)}_{E}$
in Appendix B.
It should be emphasized that, by substituting ordinary kernels into propagators,
each diagram in fact contains an infinite number of loop contributions at every 
substituting position, as shown in Fig.(\ref{fig:GA1_demo}).

Numerically, the Gaussian damping of the multipoint matter propagator $\Gamma_m^{(n)}$ 
will eventually contribute to the improvement of the convergence in calculating nonlinear 
matter power spectrum. 
Therefore, it is important to check whether $\Gamma_E^{(n)}$ share the similar 
feature.
From diagrams representing $\Gamma_E^{(1)}(k)$, we see that 
there exists a straight path through the diagram carrying the
same momentum $\vk$ at both the start and end. Along this path, there is one
convolution bias vertex $b_n$, as well as one $(n+1)$-point propagator $\Gamma_{m}^{(n)}$.
For $n=1$ (e.g.\ the third diagram in Fig.(\ref{fig:GA1_1lp})), this contribution 
recovers the same $k$-dependence as $\Gamma_{m}^{(1)}(k)$, rescaled by a constant from loop 
integration.  When $n>1$, Eq.(\ref{eqn:gammpsi_apprx}) suggests a similar 
damping of $\vk$ given $\vp_1 \cdots \vp_{n-1}$. 
Meanwhile every propagator associates with a smoothing window function $W(kR)$.
Therefore one should expect that, in the large-$k$ limit, $\Gamma_{E}^{(1)}$ will
decay as a combined effect of a Gaussian damping of $\Gamma_{m}^{(1)}$ and the smoothing
window function.

Besides the tree level result of two-point propagator $\Gamma_E^{(1,~{\rm tree})}(k)$,
which equals the one of dark matter field times the linear bias coefficient 
$b_1(k) \Gamma_m^{(1)}(k)$, 
the loop integration from higher-order calculation, e.g.\ the third diagram in 
Fig.(\ref{fig:GA1_1lp}), will also change the small-$k$ normalization
of $\Gamma_E^{(1)}(k)$ and eventually the effective large-scale bias of the
power spectrum $b^{\ast}_1$. 
As we will see at the end of this section, $\Gamma_E^{(1)}(k)$ dominates the
contribution of $b^{\ast}_1$, hence the accuracy of the 
$b^{\ast}_1$ depends on the precision of estimating the $\Gamma_E^{(1)}(k)$. 
Due to the exponential decay of matter propagators $\Gamma_m^{(n)}$, 
our resummed formula will help us in improving the convergence of 
$b^{\ast}_1$.

\begin{figure}[!htp]
\begin{center}
\includegraphics[width=0.48\textwidth]{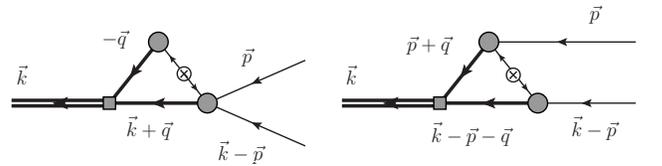}
\end{center}
\caption{\label{fig:G2_example} Examples of two different contributions of 
three-point propagator, full set of diagrams is shown in the Appendix B.
}
\end{figure}

Now, we can further derive the three-point propagator similarly, except that
two distinct contributions have to be taken into account because of the 
second order functional derivative in the definition of $\Gamma^{(2)}_{E}$. 
\begin{widetext}
\begin{eqnarray}
\label{eqn:G2_n}
\Gamma^{(2,~n)}_{E,~ab} (\vk_1, \vk_2;\eta)  
&=&  \frac{n}{2} ~ \int d^3\vq_{1\cdots n} ~ 
\delta_D(\vk-\vq_{1\cdots n}) ~\frac{b_n(\vq_1,\cdots,\vq_n)}{n!}
 \biggl[ \left \langle 
\frac{\delta^2 \psid(\vq_1, \eta)} {\delta \phi_a \delta \phi_b} 
\psid(\vq_2, \eta) \cdots \psid(\vq_n, \eta) \right \rangle 
\nonumber \\
 && + ~(n-1) ~\left\langle  \frac{\delta \psid(\vq_1, \eta)} 
 {\delta \phi_a } \frac{\delta \psid(\vq_2, \eta)} 
 {\delta \phi_b } \psid(\vq_3, \eta) \cdots \psid(\vq_n, \eta) \right\rangle 
\biggr ] 
\end{eqnarray}
\end{widetext}
As shown in Eq.(\ref{eqn:G2_n}), the first term takes the joint average 
of the products of a second derivative with $n-1$ density fields.
Diagrammatically speaking, it corresponds to selecting two branches
from a single $\delta_m$, i.e. the first diagram in Fig.\ (\ref{fig:G2_example}). 
In  Fig.\ (\ref{fig:GA2_1lp}) of Appendix C, this category includes the first, 
third and sixth  diagram.
The rest of the diagrams then correspond to the second term of Eq.(\ref{eqn:G2_n}),
where two branches originate from two different density fields.
Both terms can be written in the same form of Eq.(\ref{eqn:G1n}),
\begin{widetext}
\begin{eqnarray}
\label{eqn:G2_n_detail}
\Gamma^{(2,~n)}_{E,~ab} (\vk_1, \vk_2;\eta)  
&=&  \sum_{\{t_{ij}\}} ~ \int ~ \biggl[ 
\prod_{ 1\leq i<j\leq n} 
d^3 \vp^{ij}_{1\cdots t_{ij}} \biggr] \frac{b_n(\vk,\vp^{ij})}{n!}~
\biggl [ g^{(2,~n)}_{\{t_{ij}\}} ~
\Gamma_{m,~ ab}^{(t_1 +2)} (\vk_1,\vk_2;\vp^{11}_{1\cdots t_{11}},
\cdots, \vp^{1n}_{1\cdots t_{1n}}) \nonumber \\
&& \times \cdots \times ~ \Gamma_{m }^{(t_n)} (\vp^{n1}_{1\cdots t_{n1}},\cdots,
\vp^{nn}_{1\cdots t_{nn}} ) ~
  + ~\tilde{g}^{(2,~n)}_{\{t_{ij}\}} ~
\Gamma_{m,~ a}^{( t_1+1)} (\vk_1;\vp^{11}_{1\cdots 
t_{11}}, \cdots, \vp^{1n}_{1\cdots t_{1n}}) \nonumber \\
&& \times ~ \Gamma_{m,~ b}^{( t_2+1)} (\vk_2;\vp^{21}_{1\cdots 
t_{21}}, \cdots, \vp^{2n}_{1\cdots t_{2n}})
\cdots  \Gamma_{m}^{(t_n)} (\vp^{n1}_{1\cdots t_{n1}},\cdots,
\vp^{nn}_{1\cdots t_{nn}} ) \biggr ] 
\prod_{1\leq i<j\leq n} \left[ P_0(p^{ij})  \right]^{t_{ij}} 
\nonumber \\
\end{eqnarray}
\end{widetext}

The difference between the two terms can be clearly seen from their orders of the 
matter propagators.  With the same labeling system, the first gives $\Gamma^{(t_1+2)}_{m} 
\Gamma^{(t_2)}_{m} \cdots \Gamma^{(t_n)}_{m}$, while the second
gives $\Gamma^{(t_1+1)}_{m} \Gamma^{(t_2+1)}_{m} \cdots 
\Gamma^{(t_n)}_{m}$.
Meanwhile, the two $g$ coefficients equal
\begin{eqnarray}
 g^{(2,~n)}_{\{t_{ij}\}}& =&  \frac{n}{2} (t_1 +2)(t_1 +1) ~
\biggl[ \prod_{1\leq i\leq n} ~ {t_i \choose t_{i1}\cdots t_{in} }\biggr] \nonumber\\
 && \times ~\biggl[\prod_{1\leq i<j\leq n} (t_{ij})! \biggr], \nonumber \\
 \tilde{g}^{(2,~n)}_{\{t_{ij}\}}& =& \frac{n(n-1)}{2} (t_1+1)
 (t_2 +1) ~ \biggl[ \prod_{1\leq i\leq n} ~ {t_i \choose t_{i1}\cdots t_{in} }\biggr] 
 \nonumber\\
 &&\times~\biggl[\prod_{1\leq i<j\leq n} (t_{ij})! \biggr].
\end{eqnarray}
The first contribution $g^{(2,~n)}_{\{t_{ij}\}}$ gives $n (t_1+2)(t_1+1)$, 
while the second term $ \tilde{g}^{(2,~n)}_{\{t_{ij}\}}$ gives $ n(n-1)(t_1+1)(t_2+1)$.

At tree level, there are two diagrams: one is
$b_1(k)\Gamma^{(2)}_{m}$, from the single-$\delta_m$ contribution;
and the other is the two-$\delta_m$ term with $t_{12}=0$ for the second term of 
Eq.(\ref{eqn:G2_n_detail}),
\begin{eqnarray}
\Gamma^{(2,~{\rm tree})}_{E,~ab} (\vp,\vk-\vp; \eta; R) = b_1(\vk)~
 \Gamma_{m,~ab}^{(2)}(\vp,\vk-\vp;\eta; R) \nonumber \\
  + ~\frac{b_2(\vk-\vp,\vp)}{2} ~\Gamma_{m,~a}^{(1)}(\vp; \eta; R) 
\Gamma_{m,~b}^{(1)}(\vk-\vp; \eta; R). \quad  \nonumber \\ 
\end{eqnarray}
All the one-loop contributions can be found in the Appendix C.

Finally, the non-linear power spectrum of the field $\delta_E$ can be expressed as 
\begin{eqnarray}
\label{eqn:P_phi}
P_E(k;\eta) &=& \sum_{n \ge 1} n! \int d^3\vq_{1 \cdots n} ~ 
\delta_D(\vk - \vq_{1 \cdots n}) 
 P_0(q_1) \cdots P_0(q_n) \nonumber \\
 && \times ~~ \left [ \Gamma_E^{(n)}(\vq_1, \cdots, \vq_n; \eta)  \right ]^2 
\end{eqnarray}

\begin{figure}[!htp]
\begin{center}
\includegraphics[width=0.5\textwidth]{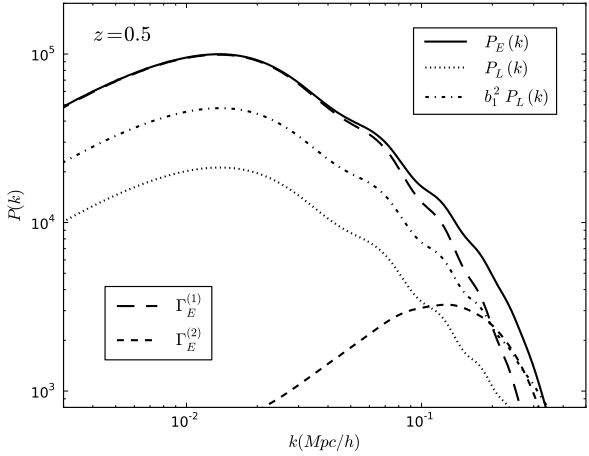}
\end{center}
\caption{\label{fig:pek} The power spectrum of Eulerian bias model
at $z=0.5$ with $b_1=1.5,~b_2=0.5$ and $b_3 = 0.2$, assuming a spherical top-hat 
smoothing window function with $R=2Mpc/h$.
The solid line gives the full result $P_E(k)$ including both
 $\Gamma^{(1)}_E$ and $\Gamma^{(2)}_E$ contributions, while each of them are represented 
 in long dashed as well as short dashed line respectively. 
 Here, $\Gamma_E^{(1)}$ is calculated up to two-loop order and $\Gamma_E^{(2)}$ to the
  one-loop order.
The dotted line shows the linear matter power spectrum $P_L(k)$, and the dot-dashed line
illustrates the linear bias contribution $b_1^2 P_L(k)$. The offset between the
dot-dashed and solid curve indicates that 
higher order bias parameters contribute to the large-scale amplitude of the power spectrum. 
}
\end{figure}

\begin{figure*}[!htp]
\begin{center}
\includegraphics[width=0.49\textwidth]{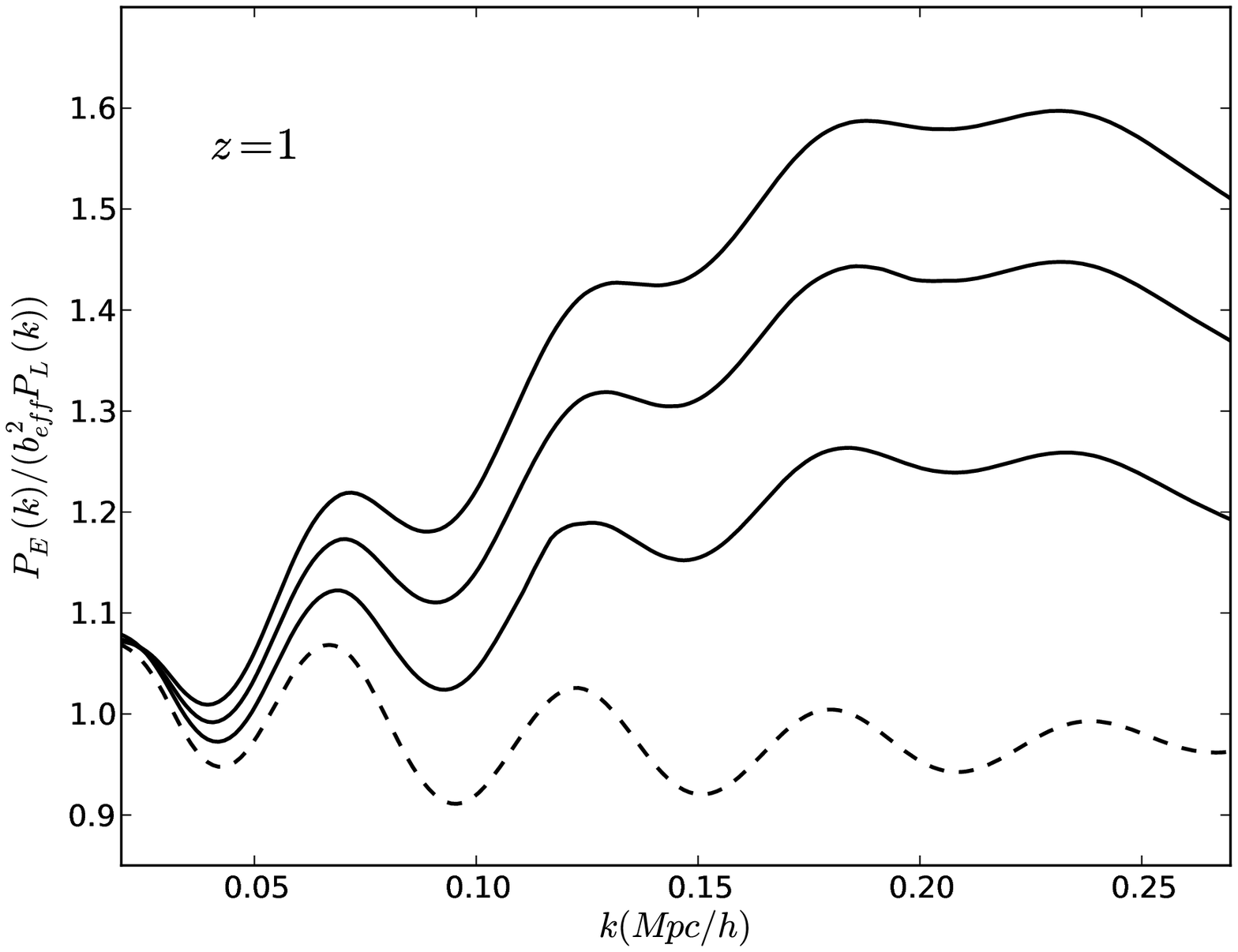}
\includegraphics[width=0.49\textwidth]{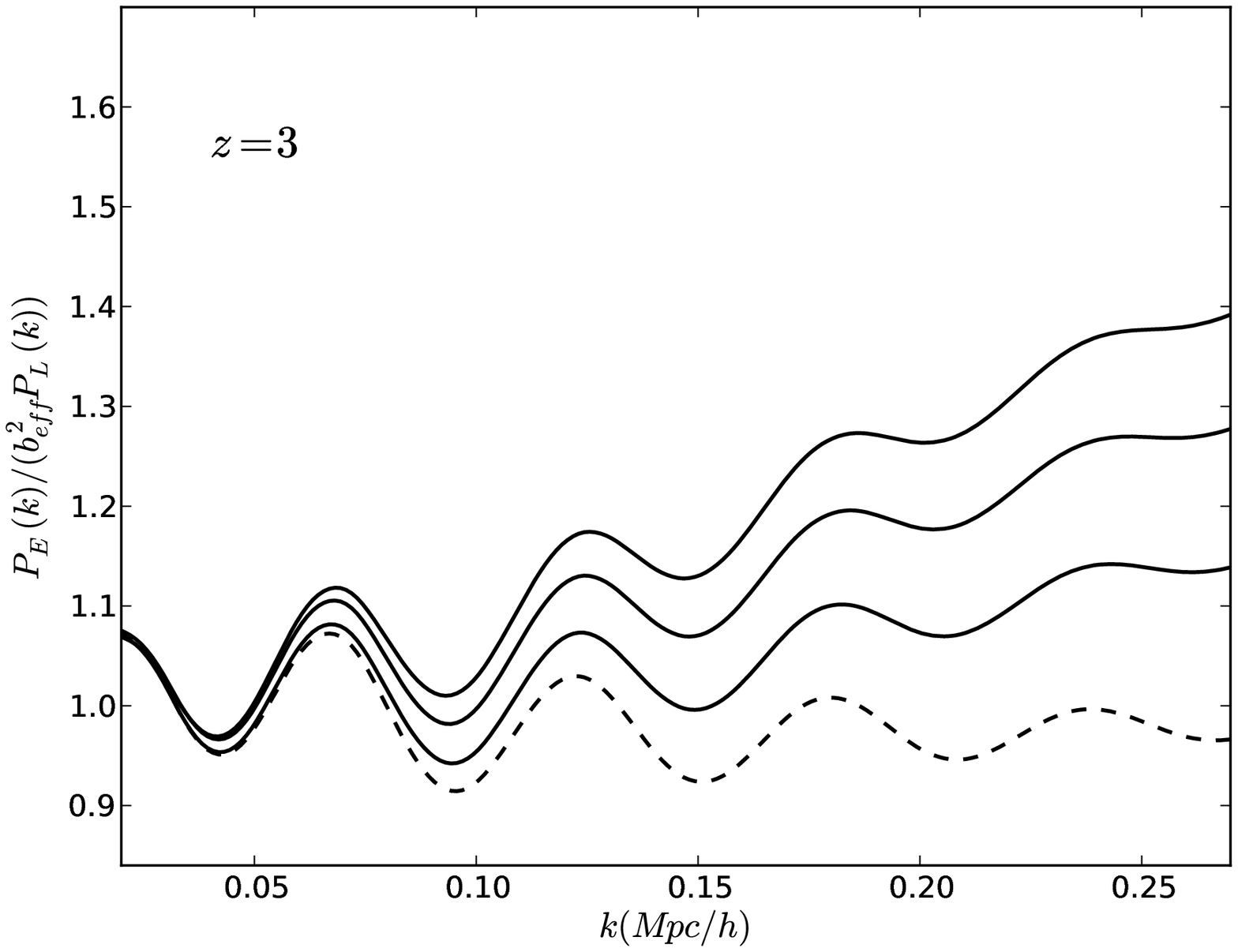}
\end{center}
\caption{\label{fig:bao} Examples of scale-dependent deviation of nonlinear Eulerian
bias model at scales of baryonic acoustic oscillation for various bias parameters at redshift $z=1$ ($\it left$) and $z=3$ ($\it right$). 
Both $\Gamma_E^{(1)}$ and $\Gamma_E^{(2)}$ are calculated up to the one-loop order. 
From the top to bottom, we adopt bias parameters: (1) $b_1=0.8,~b_2=1.5,~b_3=0.4$; 
(2) $b_1=0.8,~b_2=1,~b_3=0.2$; (3) $b_1=1,~b_2=0.5,~b_3=0.2$.
}
\end{figure*}

In Fig. (\ref{fig:pek}), we illustrate the numerical calculation of the power spectrum
of Eulerian bias model at redshift $z=0.5$ for bias parameters 
$b_1=1.5,~b_2=0.5$ and $b_3=0.2$, assuming a  spherical top-hat smoothing 
length $R=2\Mpc/h$. The solid line gives the full nonlinear power spectrum of $P_E(k)$ 
including both $\Gamma^{(1)}_E$ and $\Gamma^{(2)}_E$ contributions. 
Each of them is then represented by the long dashed and the short dashed line respectively. 
We only calculate the $\Gamma^{(2)}_E$ up to one-loop order while $\Gamma_E^{(1)}$ to
the two-loop order.
Compared to the dot-dashed line representing the linear bias contribution 
$b_1^2 P_L(k)$, one sees clearly that higher order bias parameters contribute to the
large-scale amplitude of the power spectrum.
This can also be seen in the left panel of Fig.\ (\ref{fig:pik_z3}), where we 
plot the normalized two-point propagator $\Gamma_E^{(1)}(k,z)/D(z)$ for the same 
bias model but at various redshifts, starting from $z=6$ to $z=0.5$.
At lower redshift, the variance of the local density field grows larger, so does
the effective linear bias $b^{\ast}_1$ as shown in the figure.
For the same reason, the exponential damping length decrease towards higher redshift.
As expected, all of lines at large scale lie above the true linear bias $b_1$, 
i.e. the dotted  horizontal line. 
We want to remind here that the accuracy of such effective large scale bias calculated
by our formalism has already been verified with the simulation in \cite{W11} for a
logarithmic transformation. It is guaranteed by incorporating the matter propagator, 
since the loop integrations, that contribute to $b^{\ast}_1$, are reduced and therefore
the series expansion is regulated.

Another distinguishing feature provided by our formula is the scale-dependent bias.
In Fig.(\ref{fig:bao}), we give several examples of such scale-dependent deviation at 
baryonic acoustic oscillation sclaes compared to the linear matter power spectrum at two
different redshifts for various bias parameters. They are calculated by dividing out
the no-wiggle power spectrum \citep{EH98} with appropriate normalizations.
As can be seen in the figure, such scale dependence is, roughly speaking, determined by
the relative value of linear bias $b_1$ and high-order $b_n$. So for a fixed linear bias, 
a bigger high-order $b_n$ leads to a larger scale-dependent deviation, while a smaller
$b_1$ results in the similar trend if $b_n$ are fixed. 
Not surprisingly, the same bias model will give a larger deviation at lower redshift.


\begin{figure}[!htp]
\begin{center}
\includegraphics[width=0.4\textwidth]{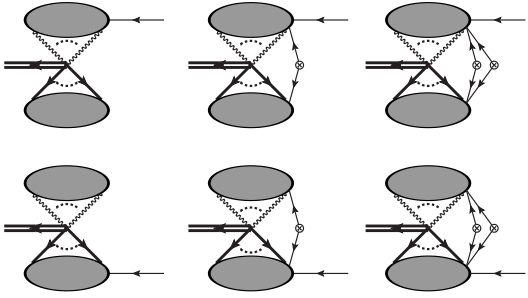}
\end{center}
\caption{\label{fig:LPT_bg}  Diagrams of $\Gamma_g^{(1)}(\vk)$ up to the two-loop
order. The first row represents the contribution of $\Gamma_{g,~2}^{(1)}(k)$ and the 
second row for $\Gamma_{g,~1}^{(1)}(k)$ terms.
}
\end{figure}

\begin{figure*}[!htp]
\begin{center}
\includegraphics[width=0.49\textwidth]{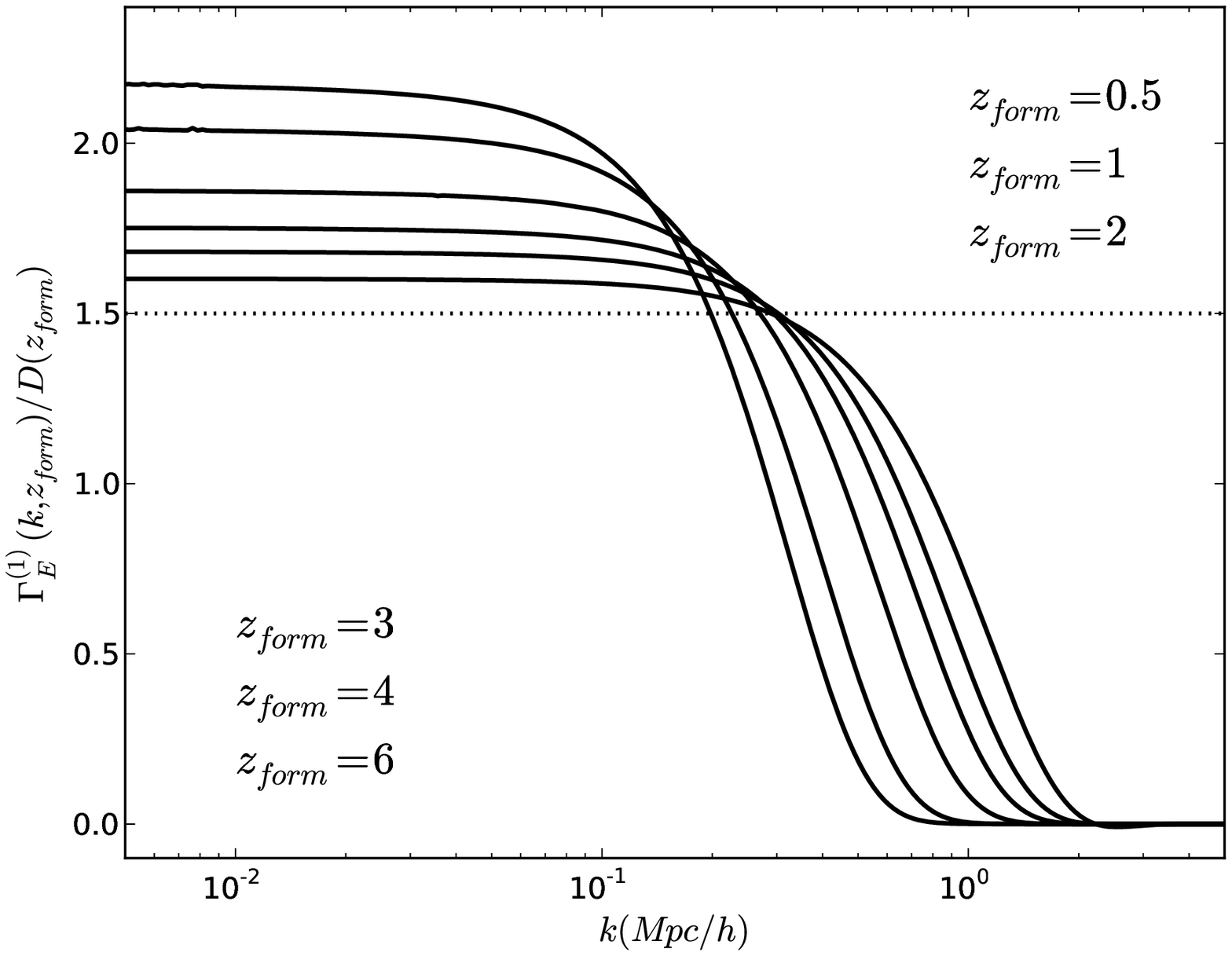}
\includegraphics[width=0.49\textwidth]{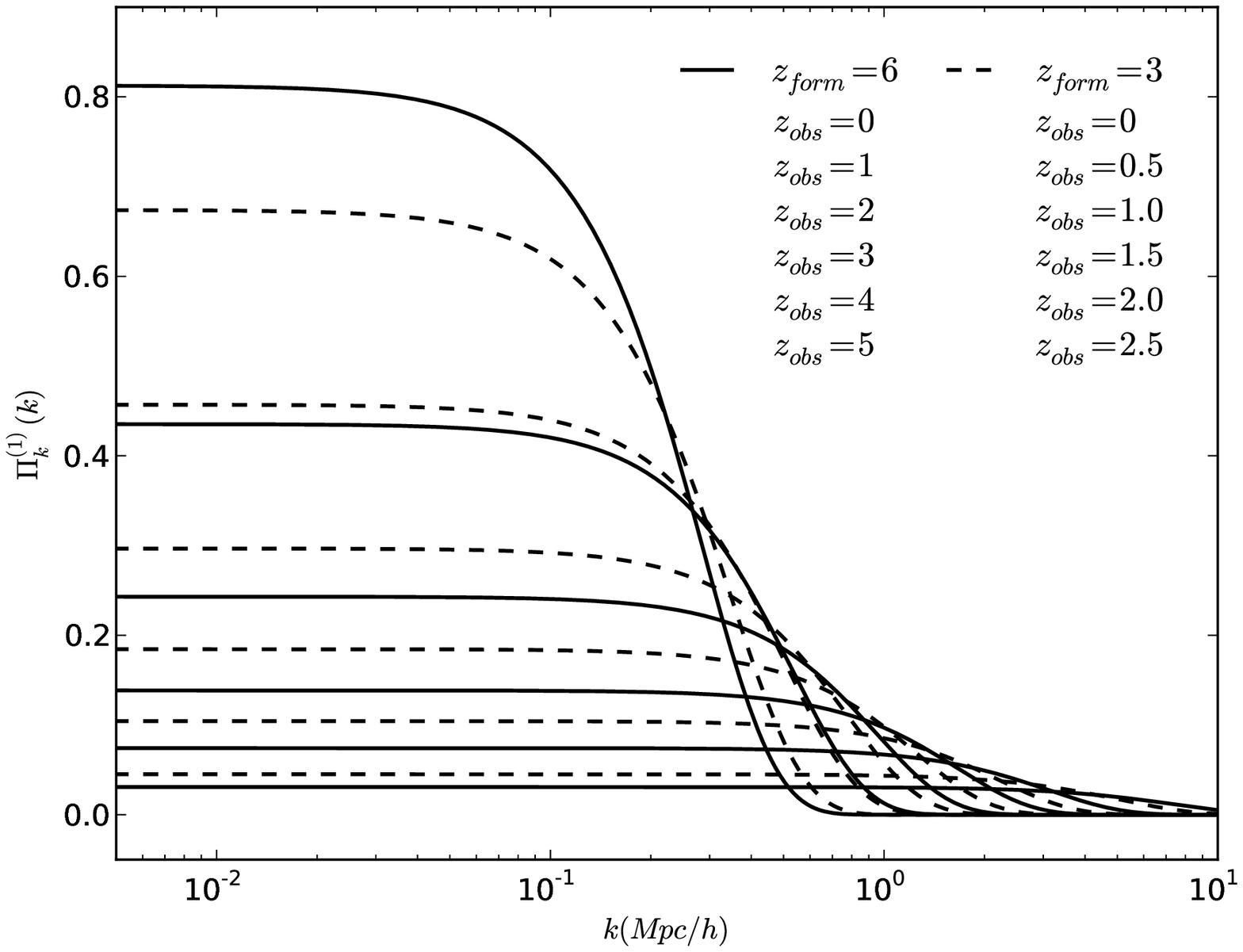}
\end{center}
\caption{\label{fig:pik_z3} 
({\it left}): Two-point propagator of Eulerian bias model $\Gamma_E^{(1)}(k)$
at various redshifts $z_{\it form} = 0.5,~1,~2,~3,~4$ and $6$ from top to bottom.
We assume the bias parameters $b_1=1.5,~b_2=0.5,~b_3=0.2$ and $0$ otherwise.
({\it right}): $\Pi_k^{(1)}(\vk;z_{\it form}, z_{\it obs})$ (Eq.\ \ref{eqn:cong_pkn}), 
from $z_{\it form} = 3$ and $6$ evolved  to $z_{\it obs}=0, ~0.5, ~1, ~ 1.5,~ 2$ 
and $2.5$ ({\it top to bottom}).
}
\end{figure*}

\section{Resummed Perturbation Theory of Continuous galaxy formation Model}
Based on the construction of resummed perturbation theory of Eulerian bias
model in the last section, we will further extend such model to incorporate
a continuous galaxies formation history. 
As shown in Eq.(\ref{eqn:deltag_c}), the perturbation of galaxy number density 
observed at $\eta_0$ can be expressed as, 
\begin{eqnarray}
    \label{eq:deltag_EL}
    \delta_g(\vk,\eta_0)&=&\Xi_g(\vk,\eta_0) 
    = \int_{\eta_{min}}^{\eta_0} d\eta ~f(\eta)\Delta_g(\vk,\eta,\eta_0). \nonumber\\
\end{eqnarray}
where the integrand $\Delta_g(\vk;\eta,\eta_0)$ is the density contrast of galaxies
formed at $\eta$ and then been observed at $\eta_0$

\begin{eqnarray}
    \label{eqn:Dgk}
    \Delta_g(\vk)& =& \int d^3\vq
    ~e^{-i\vk\cdot \vq}\left[ F[\delta_m(\vq,\eta),\eta] 
    e^{-i\vk\cdot \vpsi_g(\vq,\eta,\eta_0)} -1\right] \nonumber\\
    &=& \left[ \widetilde{F} \ast
     \widetilde{e^{-i\vk\cdot \vpsi_g}}  \right]  (\vk,\eta,\eta_0)
     -\delta_D(\vk) \nonumber \\
    &=& \sum_{n+m\ge 1}^{\infty} 
     \frac{(-i)^m}{n!~m!}\int d^3\vk_{1\cdots n}d^3\vk^{\pr}_{1\cdots m} \nonumber \\
         &&\times ~ \delta_D(\vk-\vk_{1\cdots n}-\vk^{\pr}_{1\cdots m}) 
         b_n(\vk_1,\cdots,\vk_n;\eta) \nonumber\\
         &&\times ~\left[ \delta_m(\vk_1,\eta)\cdots  \delta_m(\vk_n,\eta)\right]
         ~ \left[\vk \cdot \vpsi_g(\vk^{\pr}_1,\eta,\eta_0)\right] \nonumber \\
      &&\times ~ \cdots~ \left[\vk \cdot \vpsi_g(\vk^{\pr}_m,\eta,\eta_0)\right]
\end{eqnarray}
where $\widetilde{F}$ and $\widetilde{e^{-i\vk\cdot \vpsi_g}}$ are Fourier transforms
of $F$ and $e^{-i\vk\cdot \vpsi_g}$ respectively, and $\ast$ denotes the convolution.
Apart from the weighted average of $\Delta_g$ with the galaxy formation rate $f(\eta)$, 
this model differs 
from the work of \cite{M08b} in two ways. First, the non-linear bias functional $F[\delta_m]$ depends on the nonlinear evolved matter field $\delta_m(\eta)$ instead 
of initial state $\delta_0$. Secondly, the displacement field $\vpsi_g(\eta,\eta_0)$ 
characterizes the movement of the galaxy from location $\vq$ at time $\eta$ to Eulerian
position $\vx$ of time $\eta_0$. 
In the following, we will construct the propagator of $\delta_g(\vk,\eta_0)$ instead of 
following the work of \cite{M08b} exactly.

Starting from Eq.(\ref{eq:deltag_EL}), one can define the multi-point propagator 
of galaxies similarly. For the two-point propagator $\Gamma_g^{(1)}(k)$, we have
\begin{eqnarray}
    \left\langle  \frac{\delta \Xi_g(\vk,\eta_0)}{\delta \phi(\vk^{\pr})}
     \right \rangle  &=&  \Gamma^{(1)}_g (\vk,\eta_0) \delta_D(\vk-\vk^{\pr}) \nonumber\\
     &=& \int_{\eta_{min}}^{\eta_0}d\eta~f(\eta)~
     \left\langle \frac{\delta \Delta_g(\vk,\eta,\eta_0)}{\delta \phi(\vkpr)} 
     \right \rangle ~~
\end{eqnarray}
Concentrating on the quantity inside the integration, and we define
\begin{eqnarray}
    \left\langle  \frac{\delta \Delta_g(\vk;\eta,\eta_0)}{\delta \phi(\vk^{\pr})}
     \right \rangle 
     &=&  \Gamma_g^{(1)}(\vk;\eta,\eta_0) \delta_D(\vk-\vkpr) 
\end{eqnarray}
Substituting the expression of Eq.(\ref{eqn:Dgk}) into this definition, 
one immediately obtains two different contributions arising from the functional derivative
\begin{eqnarray}
    \label{eqn:gamma_bg1}
\Gamma_g^{(1)}(\vk;\eta,\eta_0) = \Gamma^{(1)}_{g,~1}(\vk;\eta,\eta_0) + 
\Gamma^{(1)}_{g,~2 } (\vk;\eta,\eta_0) \qquad \qquad \nonumber \\
 = \left \langle 
\left[  \frac{\delta}{\delta \phi(\vk)} \widetilde{F} \right] \ast
  \widetilde{e^{-i\vk\cdot \vpsi_g}}   \right \rangle  
+  \left \langle \widetilde{F}\ast
 \left[ \frac{\delta}{\delta \phi(\vk)} \widetilde{e^{-i\vk\cdot \vpsi_g}} \right]
  \right \rangle \nonumber \\
\end{eqnarray}
Concentrating on $\Gamma_{g,~1}^{(1)}(\vk;\eta,\eta_0)$ first, 
a further perturbative expansion gives
\begin{eqnarray}     
   \Gamma^{(1)}_{g,~1}  = \biggl \langle \sum_{n=1}^{\infty}\int d\vk_{1\cdots n}
      \frac{b_n(\vk,\vk_1,\cdots \vk_n)}{n!}  \biggl[
    \delta_m(\vk_1)\cdots\delta_m(\vk_n)  ~ \nonumber \\
     \times ~\frac{\delta}{\delta\phi}\delta_m(\vk) \biggr] \times 
     \sum_{m=0}^{\infty} \frac{(-i)^m}{m!}\int d\vkpr_{1\cdots m}
     \left[ \vk\cdot\vpsi_g(\vkpr_1) \right] \cdots   \nonumber \\
   \times~ \cdots ~ \left[ \vk\cdot     \vpsi_g(\vkpr_m) \right] \biggr \rangle 
    \qquad\qquad\qquad\qquad\qquad  \qquad \quad~
\end{eqnarray}
Without going into detail, one finds the resemblance between above equation
and Eq.\ (\ref{eqn:G1_2_def}, \ref{eqn:G1_2}), since both $\widetilde{F}$ and
$\widetilde{e^{-i\vk\cdot \vpsi_g}}$ can be written in the form of Eq.\ (\ref{eqn:xi_def}).
Therefore, following the exact same procedure after Eq.\ (\ref{eqn:G1_2_def}), 
the contribution $\Gamma^{(1)}_{g,~1}$ can be similarly resummed using the
 $\Gamma-$expansion, and one simply reads from Eq.\ (\ref{eqn:G1_2_final}) as
\begin{eqnarray}
    \label{eqn:gamma_bg1_1}
    \Gamma^{(1)}_{g~,1}(\vk;\eta,\eta_0) = \sum_n (n+1)!
   \int d^3\vp_{1\cdots n} ~  \biggl[P_0(\vp_{1\cdots n})\biggr]^n  \nonumber\\
   \times ~~ \Gamma_E ^{(n+1)}  (\vk,\vp_1,\cdots,\vp_n;\eta) 
   \Pi_k^{(n)} (\vp_1,\cdots,\vp_n;\eta,\eta_0)
  \quad
\end{eqnarray}
Here the first contribution, arisen from $\langle \delta \tilde{F} / \delta \phi(\vk) 
  \rangle$, is the multipoint propagator of Eulerian biased tracer 
$\Gamma_E^{(n)}$ which we have discussed in the last section, and the second 
contribution is defined as
\begin{eqnarray}
    \label{eqn:cong_pkn}
    \Pi_k^{(n)}(\vp_1,\cdots,\vp_n;\eta,\eta_0)= \frac{1}{n!}\left \langle 
    \frac{\delta^n ~\widetilde{e^{-i\vk\cdot \vpsi_g } } (\vp;\eta,\eta_0)}
    {\delta \phi(\vp_1) \cdots \delta \phi(\vp_n)}   \right \rangle \quad
\end{eqnarray}
where $\vp=\vp_{1\cdots n}$.  When $\vk =\vp$, it recovers to the 
definition of multipoint matter propagators in Lagrangian perturbation theory
with galaxy displacement field $\vpsi_g(\eta,\eta_0)$.

The physical consequence of Eq.\ (\ref{eqn:gamma_bg1_1}) is the separation between
the nonlinear gravitational evolution of galaxies after their creation characterized 
by $\Pi_k^{(n)}$ and the more complicated nonlinear galaxy formation physics 
parameterized by the unknown functional $F[\delta_m]$. The final nonlinear growth of 
galaxy density perturbation $\Gamma_g^{(1)}$ relates to these two physical processes 
in a statistical way. 
Diagrammatically, Starting from the second diagram in Fig.\ (\ref{fig:LPT_bg_ele}),
one can construct $\Gamma_E^{(n)}$ and $\Pi_k^{(n)}$ simply by gluing initial states
connected to wavy line and thick solid line separately. 
We illustrate $\Gamma_{g,~1}^{(1)}(k)$ in the second row
of Fig. (\ref{fig:LPT_bg}) up to the two-loop order.

Similarly, the second term in Eq.\ (\ref{eqn:gamma_bg1}) 
\begin{eqnarray}
    \Gamma^{(1)}_{g,2}
     = \biggl \langle     \sum_{n=1}^{\infty}\int d\vk_{1\cdots n}
      \frac{b_n(\vk_1,\cdots \vk_n;\eta)}{n!} 
    \delta_m(\vk_1)\cdots\delta_m(\vk_n)  \nonumber \\
      \times \sum_{m=0}^{\infty} \frac{(-i)^m}{m!}\int d\vkpr_{1\cdots m}
     \left[ \vk\cdot\vpsi_g(\vkpr_1) \right] \cdots  
    \left[ \vk\cdot\vpsi_g(\vkpr_m) \right]   \nonumber \\
     \times \left[ \vk\cdot\frac{\delta \vpsi_g(\vkpr_{m+1}) }
    {\delta \phi(\vkpr)} \right]  \biggr \rangle \qquad\qquad\qquad\qquad\qquad
    \qquad\quad
\end{eqnarray}
can be resummed as 
\begin{eqnarray}
     \Gamma^{(1)}_{g~,2}(\vk;\eta,\eta_0) = \sum_n (n+1)!
    \int d^3\vp_{1\cdots n} ~ \biggl[P_0(\vp_{1\cdots n})\biggr]^n 
    \nonumber \\
    \times ~   \Gamma_E ^{(n)}  (\vk,\vp_1,\cdots,\vp_n;\eta)
 \Pi_k^{(n+1)} (\vk, \vp_1,\cdots,\vp_n;\eta,\eta_0), \quad
\end{eqnarray}
which corresponds to the first row of Fig.(\ref{fig:LPT_bg}). 
Therefore, up to the one-loop order, the two-point propagator of galaxy 
density perturbation $\Gamma_g^{(1)}(k)$ equals
\begin{eqnarray}
    \label{eqn:gamma_bg1_1lp}
    \Gamma_{g}^{(1)}(\vk;\eta,\eta_0) \approx \Gamma_E^{(1)}(\vk;\eta) +
       \Pi_k^{(1)}(\vk;\eta,\eta_0) +  2\int d^3\vp~ P_0(p) \nonumber \\
  \biggl[ \Gamma_E^{(2)}(\vk,\vp;\eta) \Pi_k^{(1)}(\vp;\eta,\eta_0)
  + \Gamma_E^{(1)}  (\vp;\eta)  \Pi_k^{(2)}(\vk,\vp;\eta,\eta_0)  \biggr]\nonumber\\
\end{eqnarray}

\begin{figure*}[!htp]
\begin{center}
\includegraphics[width=0.49 \textwidth]{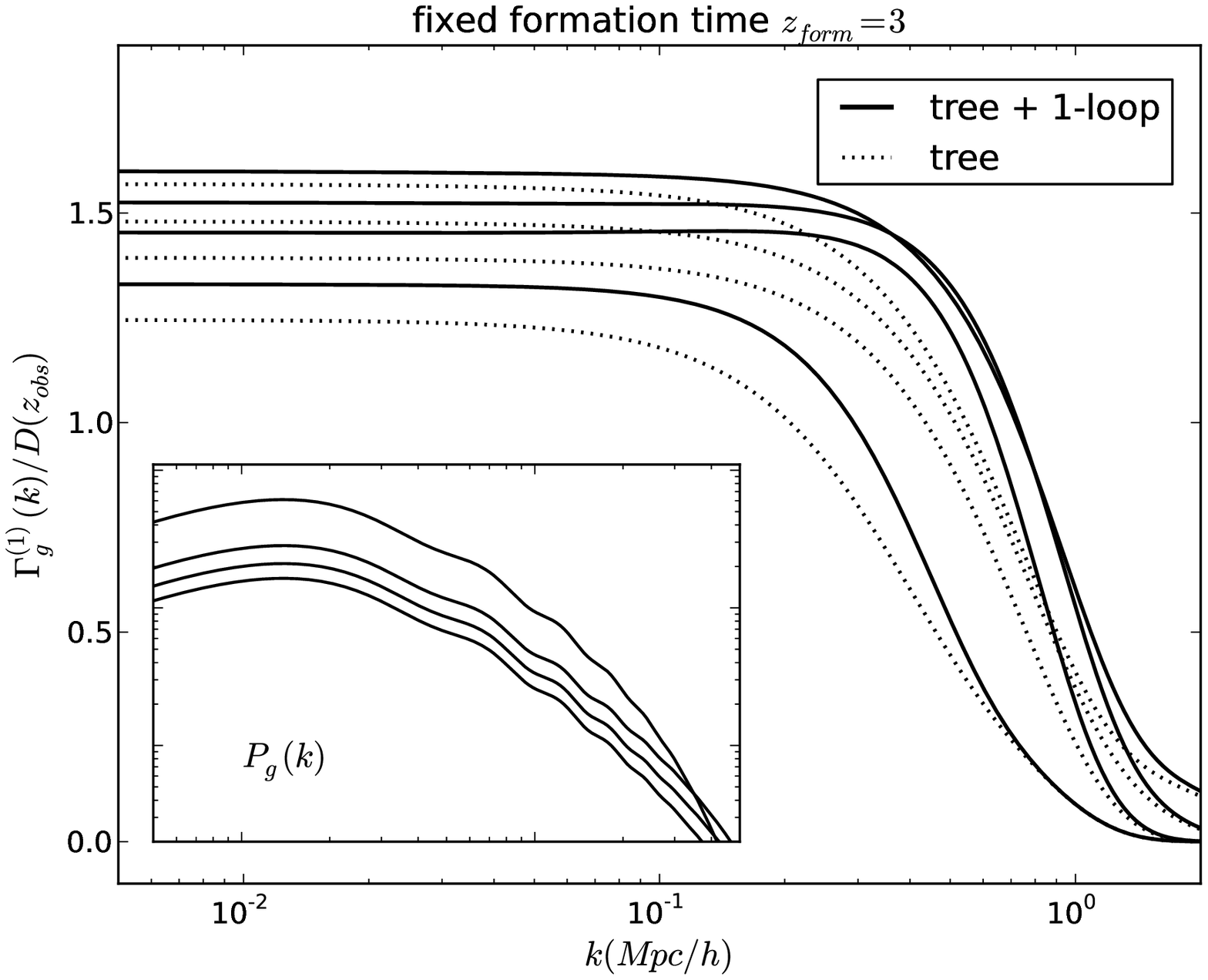}
\includegraphics[width=0.49 \textwidth]{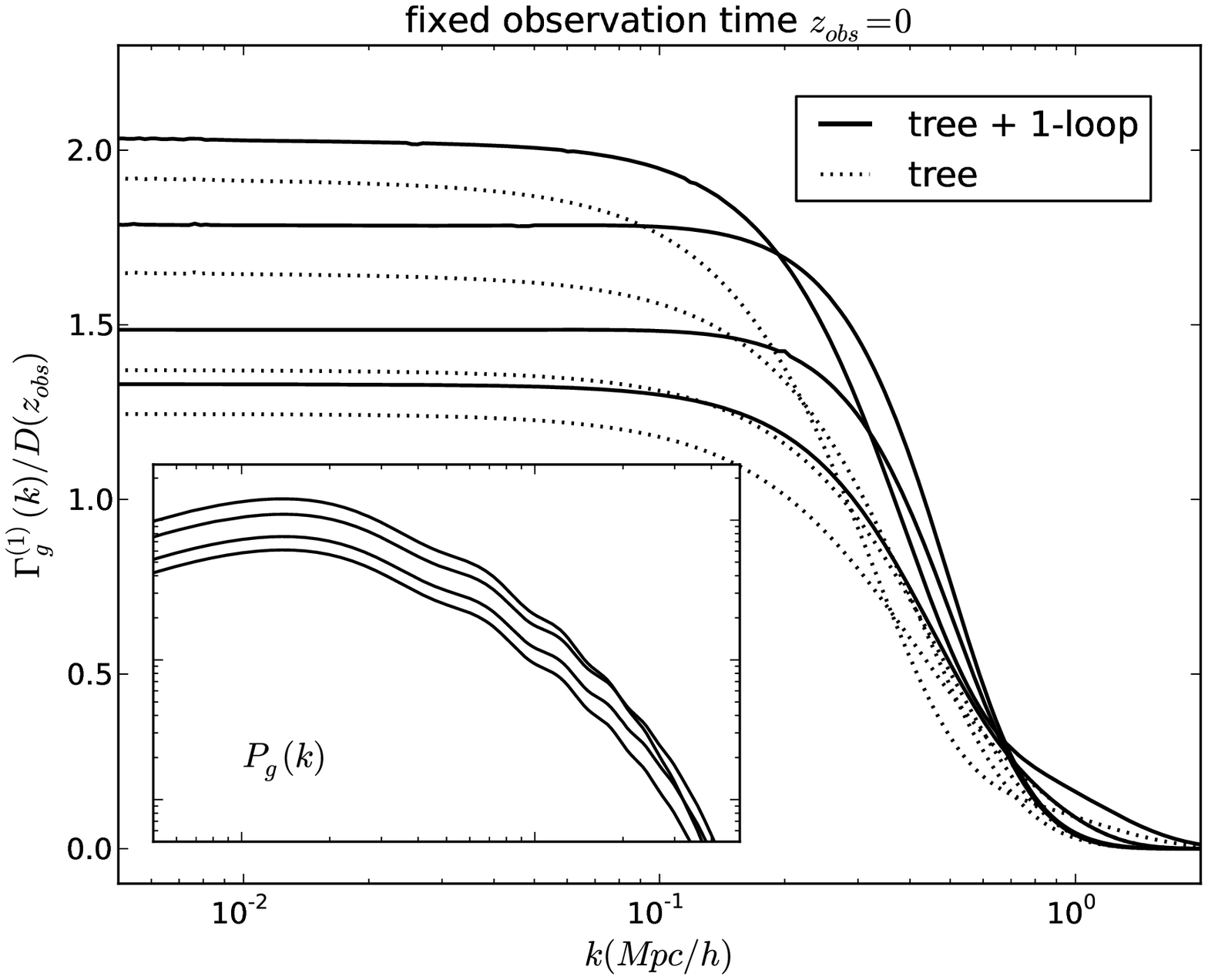}
\end{center}
\caption{\label{fig:pk_z3} The normalized two-point propagator 
$\Gamma^{(1)}_g(k;~z_{\it form},~z_{\it obs})/D(z_{\it obs})$ of galaxies which formed 
at fixed formation redshift ({\it left}) $z_{\it form}=3$, and observed at 
$z_{\it obs}=2, ~1.5,~ 1$ and $0$ ({\it top to bottom}); and 
at fixed observation redshift ({\it right}) $z_{\it obs}=0$, and formed at $z_{\it form}=0.5,~1,~2$ and $3$ ({\it top to bottom}).
Assuming nonlinear bias model $b_1 = 1.5, ~b_2=0.5, ~b_3=0.2$. 
{\it Inner~ panel:} the corresponding power spectra in the opposite/same 
({\it left}/{\it right}) order.
}
\end{figure*}

To proceed, we have to estimate the $\Pi_k^{(n)}$ with galaxy displacement field
$\vpsi_g(\eta,\eta_0)$. To the first order, the Zel'dovich approximation simply
gives
\begin{eqnarray}
    \vpsi^{(1)}_g(\vk;\eta,\eta_0) = [D(\eta_0)-D(\eta)] \delta_0(\vk),
\end{eqnarray}
and therefore, 
\begin{eqnarray}
    \label{eqn:pik_tree}
    \Pi_k^{(1)}(k;\eta,\eta_0) \approx [ D_0-D] \exp \left[ -\frac{k^2 (D_0-D)^2}{2}  \sigma_{\Psi}^2 \right],
\end{eqnarray}
where $D_0= D(\eta_0)$, and $D = D(\eta)$. So it is fully characterized by the
gravitational growth between two different moments.
Though higher order corrections analogous to the calculation in Appendix A are possible,
we haven't explicitly calculated them here. The caveat is that the galaxy displacement 
field at $\vq$ relates to $\vpsi_m$ at some initial Lagrangian position $\vq_{in}$ 
since we assume
\begin{eqnarray}
    \vpsi_g(\vq;\eta,\eta_0) = \vpsi_m(\vq_{in};\eta_0) -
     \vpsi_m(\vq_{in};\eta).
\end{eqnarray}
At the lowest order $\vq=\vq_{in}$, but the coordinate transform needs to be carefully 
incorporated for higher order calculation. 
Furthermore, a more reliable approach is to directly solve the fundamental dynamical 
equation (Eq.\ \ref{eqn:LPT_dynamics}) with the potential determined by the nonlinear
matter distribution at $\eta$. We will leave this in the future work and simply 
utilize the Zel'dovich approximation (Eq.\ \ref{eqn:pik_tree}) within this paper.

In the right panel of Fig. (\ref{fig:pik_z3}), we illustrate the $\Pi_k^{(1)}(k)$
from the moment of galaxy formation $\zform = 3$ and $6$ evolve to various
observational time $\zobs$. 
At the large scale, it converges to linear result $D(\zobs)-D(\zform)$, which
relates to the linear debiasing we discussed in Eq.\ (\ref{eqn:linear_debias}), 
and dampens to zero towards smaller scales with the speed proportional to
$[D(\zobs)-D(\zform)]^2$.
As can be seen in the figure, when galaxies have more time to evolve 
, i.e. $\zobs-\zform$ is large, the observed galaxy distribution is more 
influenced by gravitational evolution. This can be easily seen from 
Eq. (\ref{eqn:gamma_bg1_1lp}), since at tree level 
$\Gamma_g^{(1)}(k)= \Gamma_E^{(1)}(k;\zform)+\Pi_k^{(1)}(k;\zform, \zobs)$. 
When $D(\zobs)-D(\zform)$ is small, $\Pi_k^{(1)}(k;\zform,\zobs)$ vanish
and therefore the gravitational evolution is negligible. 

This can also be clearly seen in the left panel of Fig. (\ref{fig:pk_z3}), where
we plot the normalized galaxy propagator $\Gamma_g^{(1)}(k; \zform, \zobs)/D(\zobs)$
with fixed formation redshift $\zform=3$ and observed at various $\zobs$. 
The dashed line shows the tree level of Eq.\ (\ref{eqn:gamma_bg1_1lp}) and
solid line corresponds to 1-loop results. In this situation, $\Gamma_E^{(n)}$ is
fixed, and as $\zobs$ decreases from $2$ to $0$, the large scale bias also 
decreases as expected.  
However, the amplitude of the power spectrum itself as shown in the inner panel of
left Fig.\ (\ref{fig:pk_z3}), is still increasing at lower redshift due to the 
linear growth $D(\zobs)$.
From the diagrams of three-point propagator of Eulerian bias model,
at large scale $k\to 0$, $\Gamma_E^{(2)}(\vk,\vp)$ is nonzero, 
hence the loop integration also contributes to the large scale bias, although
it's still dominated by the tree-level value.
As for smaller scales, $\Gamma_g^{(1)}$ dampens qualitatively similar to
the $\Pi_k^{(1)}(k)$.

In the right panel of Fig.\ (\ref{fig:pk_z3}), we illustrate the same quantity 
for fixed observation redshift $\zobs = 0$, when galaxies are formed at different
time with the same nonlinear process (same $F[\delta_m]$ and $b_n$). 
In this case, both $\Gamma_E^{(n)}$ and $\Pi_k^{(n)}$ are changing and the
evolution of large scale bias would in principal depends on the specific parameters 
$b_n$ adopted, since the $\Gamma_E$ and $\Pi_k$ evolve with time oppositely.
In the examples we shows, $\Gamma_E^{(1)}(k)$ dominates the evolution, i.e. galaxies
formed at lower redshift would be higher biased due to the stronger nonlinear 
effects.

\begin{figure*}[!htp]
\begin{center}
\includegraphics[width=0.49 \textwidth]{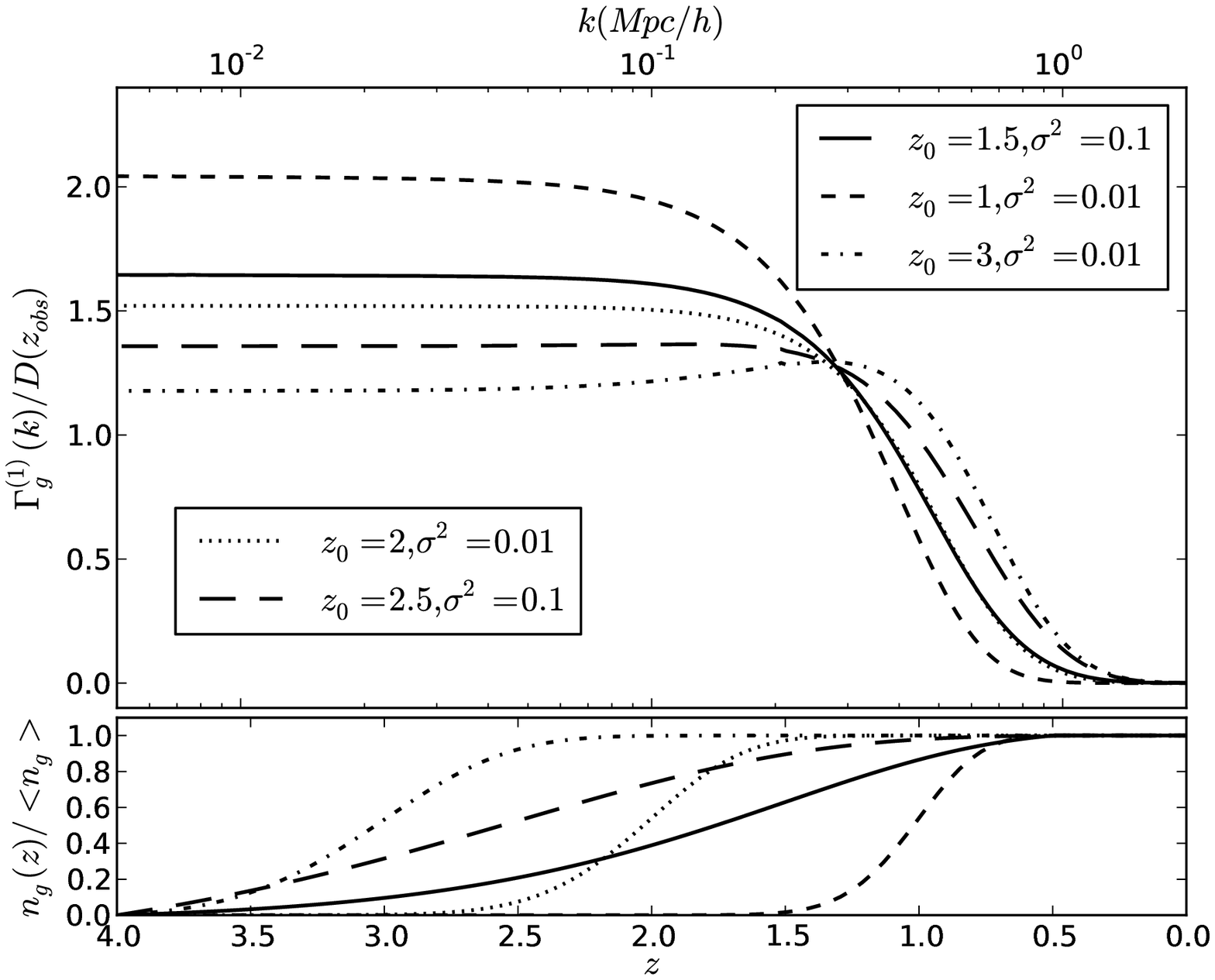}
\includegraphics[width=0.49 \textwidth]{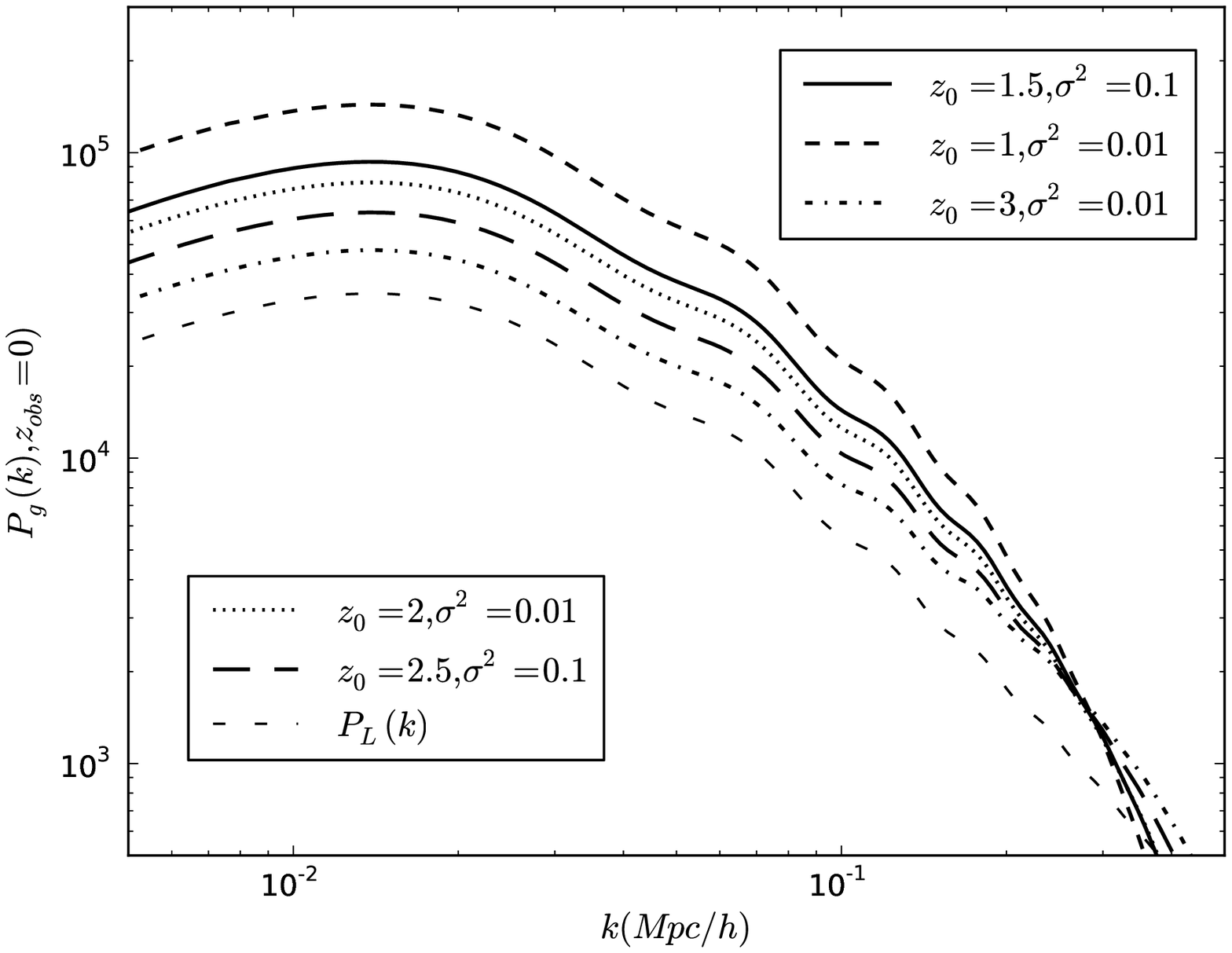}
\end{center}
\caption{\label{fig:pk_cont_zobs_0} 
The normalized two-point propagator ({\it left})
$\Gamma^{(1)}_g(k;~z_{\it form},~z_{\it obs})/D(z_{\it obs})$ and corresponding
 power spectrum ({\it right}) for various models with continuous galaxy formation.
 ({\it left-lower}): The cumulative average galaxy formation history for models shown
in the upper panel. 
}
\end{figure*}

Finally, we consider the continuous galaxy formation model, assuming a simple 
log-normal model for average galaxy formation rate \citep{CSS12}
\begin{eqnarray}
    f(z) = N_g \exp\left[ -\frac{[\log(1+z)-\log(1+z_0)]^2}{2\sigma^2} \right],
\end{eqnarray}
where $N_g$ is the normalization factor ensuring $\int dz f(z) =1$. 
Therefore, we have
two parameters characterizing the model: $z_0$ the peak redshift of galaxy 
formation, and $\sigma^2$ the width of the redshift for galaxy formation.
In the lower panel of Fig.(\ref{fig:pk_cont_zobs_0}), we give several examples
of galaxy formation history. 

In the left-upper panel of Fig.(\ref{fig:pk_cont_zobs_0}), we present again the 
normalized propagator for various galaxy formation models which we shows in the
lower panel of the same figure.
As discussed previously, the $\Gamma_g^{(1)}(k;\zobs)$ here is simply a time average
of $\Gamma_g^{(1)}(k;\zform,\zobs)$ weighted with $f(\zform)$.
Therefore, for a narrower formation history (e.g. the dashed, 
dotted and dash-dotted lines), where $\sigma^2=0.01$,
the results are similar to corresponding galaxy bursting models with different
$z_0$. On the other hand, when galaxy tends to form in a similar rate during a 
longer time (solid and long-dashed lines), e.g. $\sigma^2=0.1$, we are more likely
to observe a galaxy power spectrum with medium amplitude and nonlinear damping.
We also illustrate the corresponding power spectrum on the right panel of 
Fig.(\ref{fig:pk_cont_zobs_0}).

\section{Conclusion and Discussion}
An accurate modeling of the statistics of galaxy density perturbation is crucial
for the success of next generation galaxy surveys. 
In this paper, we considered the nonlinear bias model in the context of
resummed perturbation theory. In Section III, we discussed the Eulerian bias model
and then generalized it to incorporate the continuous galaxy formation in Section IV.
By utilizing the multipoint propagators of the matter field, our formalism for 
Eulerian bias model is more accurate than the standard approach in both the linear
as well as the quasi-linear regimes. 
This has already been verified in the work of \citep{W11}, where a good agreement
was achieved comparing to the simulation even for the slowly converging logarithmic 
mapping of the density distribution. 
However, a detailed comparison with high-precision simulation is still needed, in
at least two different levels: the accuracy of perturbative calculation itself 
compared to the cost of its numerical calculation, and the ability of describing the
scale-dependent galaxy clustering bias at the epoch of their formation. 

Furthermore, even after separating out the subsequent gravitational non-locality, 
the bias parameters could still be nonlocal. For example, the $\Delta \rho$
in Eq.\ (\ref{eqn:cont_deltarho}) only counts the effective newborn galaxies that 
finally entered into the sample, however, such description is more or less
ambiguous due to galaxy merger. Therefore, the identified progenitors in our 
formalism then include all merged galaxies, which would introduce a effective 
non-locality in the functional $F[\delta_m]$.
In principle, this could be solved by introducing another merger contributions 
in the integration, but with the cost that could further complicate the model.

\acknowledgments 
XW thanks for the productive discussion with Donghui Jeong, Mark Neyrinck, 
Patrick McDonald and Kwan Chuen Chan.
XW and AS are grateful for support from the Gordon and Betty Moore Foundations.

\appendix
\newcommand{\appsection}[1]{\let\oldthesection\thesection
 \renewcommand{\thesection}{\oldthesection}
  \section{#1}\let\thesection\oldthesection}

\appsection{Matter Density Propagator in Lagrangian Perturbation Theory}
Here we consider the generalized multipoint propagator 
\begin{eqnarray}
    \Pi_k^{(n)}(\vp_1,\cdots,\vp_n) = \frac{1}{n!}\left\langle  \frac{\delta^n 
    ~\widetilde{e^{-i\vk\cdot \vpsi}}(\vp)}{\delta\phi(\vp_1)\cdots \delta\phi(\vp_n)}
    \right\rangle
\end{eqnarray}
When $\vp_1+\cdots+\vp_n=\vk$, one retrieves the $(n+1)-$point propagator 
$\Gamma^{(n)}_m$. For $n=0$, we define $\Pi^{(0)}(\vk)$ as
\begin{eqnarray}
    \Pi^{(0)}(\vk) = \left \langle e^{-i\vk\cdot \vpsi(0)} \right \rangle
    \approx \exp\left[-\frac{k^2}{6}\langle |\vpsi(0)|^2 \rangle \right]
\end{eqnarray}
As shown in Eq.(\ref{eqn:LPT_Pi0}). 
\begin{eqnarray}
    \Pi_k^{(1)}(\vp)&=& \Pi^{(0)}(\vk)\left \langle 
    \frac{\delta }{\delta\phi(\vp)}  \widetilde{e^{-i\vk\cdot \vpsi}}(\vp)
    \right \rangle_c \nonumber\\
    &=& \Pi^{(0)}(\vk)~ T^{(1)}(\vk, \vp).
\end{eqnarray}
We then explicitly expands the terms inside ensemble average, which will be 
denoted as $T^{(1)}(\vk)$ in the following
\begin{eqnarray}
    \label{eqn:app_T1}
    T^{(1)}(\vk,\vp)& =& \sum_{n=0, m=1}^{\infty} \frac{(-i)^nD^m}{n!(m-1)!} 
    \biggl \langle \left [ 
   \int d^3\vq~ \vk \cdot \vpsi(\vq) \right]^n \int d^3\vp \nonumber\\ 
    &&  d^3\vppr_{1\cdots (m-1)} 
    \delta_D(\vp-\vk-\vppr_{1\cdots m-1}) \bigl[ \vk \cdot 
    \vL^{(m)}(\vppr_1, \nonumber \\
    && \cdots, \vppr_{m-1}, \vk)  \bigr]
     ~ \left[\delta_0(\vppr_1) \cdots \delta_0(\vppr_{m-1}) \right]  \biggr \rangle_c
\end{eqnarray}
Fig(\ref{fig:LPT_G1}) shows all the contributions up to the one-loop order. Denotes
$T^{(1)}_{n,m}(\vk)$ as individual terms been summed in above equation, we have
\begin{eqnarray}
    T^{(1)}_{0,1}&=& D(\eta) k_i L^{(1)}_i,   \nonumber \\ 
    T^{(1)}_{0,3}&=& \frac{D^3(\eta)k_i}{2} \int d^3\vq ~L_i^{(3)}(\vq,-\vq, \vp) 
    P_0(q)  \nonumber\\
    T^{(1)}_{1,2}&=& D^3(\eta) k_i k_j  \int d^3\vq ~ L^{(1)}_i(\vq)
       L_j^{(2)}(-\vq,\vp) P_0(q), 
    \nonumber \\
\end{eqnarray}
where LPT kernel $\vL^{(n)}(\vp_1,\cdots, \vp_n)$ 
\begin{eqnarray}
    \vL^{(1)} &=& \frac{\vp}{p^2} \nonumber\\
    \vL^{(2)}&=& \frac{3}{7} \frac{\vp}{p^2} \left[ 1- 
    \mu_{1,2}^2\right]  \nonumber \\
    \vL^{(3a)} &=& \frac{5}{7} \frac{\vp}{p^2} \left [ 1-\mu_{1,2}^2
     \right] [1-\mu_{12,3}^2]  -\frac{1}{3} \frac{\vp}{p^2} [1-3\mu_{1,2}^2 
     \nonumber \\
      && + 2\mu_{1,2}\mu_{2,3}\mu_{3,1} ]  + \vp \times \mathbf{T}\nonumber\\
    \vL^{(3)} &=& \frac{1}{3}[\vL^{(3a)} + {\it perm} ]
\end{eqnarray}
Substituting the definition of $\vL^{(n)}(\vp_1,\cdots, \vp_n)$ 
into Eq.(\ref{eqn:app_T1}), one can 
explicit carry out $T^{(1)}$ up to the one-loop order, 
\begin{eqnarray}
    \label{eqn:app_Tm1}
T^{(1)}_{0,1} &=& D(\eta) \frac{k_i p_i}{p^2}  \nonumber\\
T^{(1)}_{0,3} &=&  \frac{5}{21}D^3(\eta) \frac{k_i p_i}{p^2} R_1(p) \nonumber\\
T^{(1)}_{1,2} &=& \frac{3}{14} D^3(\eta)  \biggl[ \left(\frac{\vk\cdot\vp}
  {p^2} \right)^2 [R_1(p)+2R_2(p)]  \nonumber \\
   && ~~ - \frac{k^2}{p^2} R_1(p) \biggr]
\end{eqnarray}
where we have defined a slightly different version of the integral functions $R_n(k)$ 
introduced by \cite{M08a}
Here we denotes 
\begin{eqnarray}
    R_n(k) = \frac{1}{48} \frac{k^3}{4\pi^2}\int_0^{\infty} dr ~P_0(kr) \tilde{R}_n.
\end{eqnarray}
and 
\begin{eqnarray}
 \tilde{R}_1 &=& -\frac{2}{r^2}(1+r^2)(3-14r^2+3r^4) \nonumber\\
      && + \frac{3}{r^3}(r^2-1)   \ln\left| \frac{1+r}{1-r} \right|  \nonumber\\
  \tilde{R}_2 &=& \frac{2}{r^2}(1-r^2)(3-2r^2+ 3 r^4) \nonumber\\
  &&~~ + ~ \frac{3}{r^3}(r^2-1)^3(1+r^2)  \ln\left| \frac{1+r}{1-r} \right|
\end{eqnarray}
Setting $\vk = \vp$ in Eq. (\ref{eqn:app_Tm1}), we get the $T^{(1)}(k)$ up to one-loop order
\begin{eqnarray}    
    T^{(1)}_{\rm tree} &=& D(\eta) \nonumber \\
    T_{\rm 1-loop}^{(1)}(k) &=& D(\eta)^3 \int  \frac{d^3\vp}{504 k^3 p^5} P_L(p) \biggl [
    6 k^7p + 5k^5p^3 + 50  k^3p^5   \nonumber \\ &&~   -21 k p^7
    + \frac{3}{2}(k^2-p^2)^3(2k^2+7p^2) \ln\left| \frac{k-p}{k+p} \right|   \biggr]. 
    \nonumber\\
\end{eqnarray}
which coincide with the calculation of \cite{B11}.
For three-point $\Pi_k^{(2)}(\vp_1,\vp_2)$
\begin{eqnarray}
    \Pi_k^{(2)}(\vp_1,\vp_2) &=& \frac{1}{2} \Pi^{(0)}(k) 
   \left \langle \frac{\delta^2 }{\delta\phi(\vp_1)\delta\phi(\vp_2)} 
    \widetilde{e^{-i\vk\cdot \vpsi}}(\vp)  \right \rangle_c \nonumber \\
    &\approx&  \Pi^{(0)}(k) \frac{D^2(\eta)}{2}  \biggl[ k_i L^{(2)}_i(\vp_1,\vp_2) +  
    \nonumber \\
    &&~~ + ~~ k_i k_j L^{(1)}_i (\vp_1)L^{(1)}_i (\vp_2)  \biggr]
\end{eqnarray}
Assuming $\vk=\vp_1+\vp_2$, One recover the tree-level result
$\Gamma^{(2)}_m(\vk)$, 
\begin{eqnarray}
    \Gamma^{(2)}_m(\vk_1,\vk_2) = \Pi^{(0)}(\vk) F_2(\vk_1,\vk_2),
\end{eqnarray}
where $F_2$ is the second order Eulerian perturbation kernel.

\appsection{Two-loop Order of $\Gamma^{(1)}_E$}
\begin{figure*}[!htp]
\begin{center}
\includegraphics[width=0.75\textwidth]{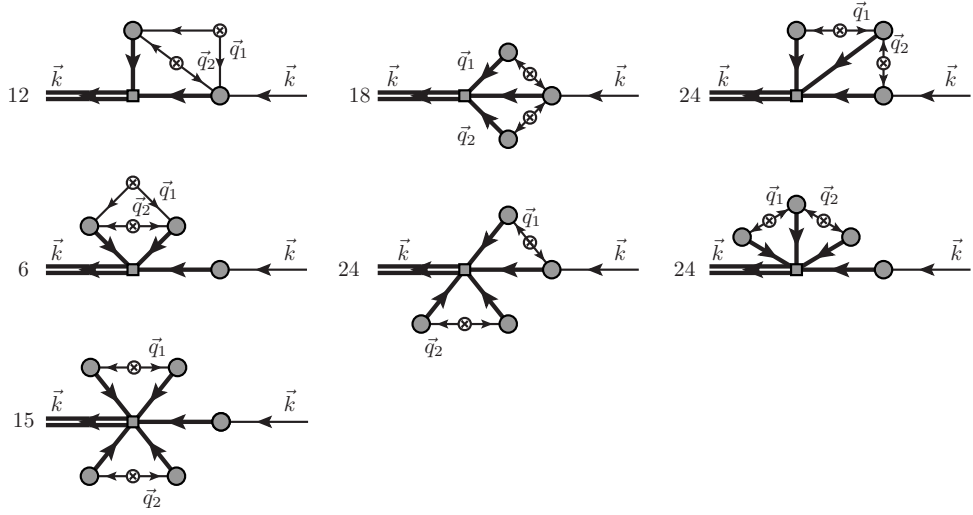}
\end{center}
\caption{\label{fig:GA1_2lp} Two-loop order of the two-point nonlinear propagator. 
}
\end{figure*}

In this paper, we have calculated the two-point propagator $\Gamma_{E}^{(1)}$ up to 
two-loop order. Seven non-vanishing contributions are depicted in 
Fig.(\ref{fig:GA1_2lp}).
From these diagrams, one can write down all terms explicitly
\begin{eqnarray}
  \Gamma_{E,~a}^{(\rm 1,~ 2lp)}(\vk; \eta) 
  = \int d^3\vp_{12} ~ P_0(p_1)P_0(p_2) ~ 
  \left[ \sum_{i=1}^{7} ~  \mK^{(1,~\rm 2lp)}_i \right ]\nonumber\\
\end{eqnarray}
where 
\begin{eqnarray}
 \mK^{(1,~\rm 2lp)}_1  &=& 6~ b_2(\vk+\vp_1+\vp_2, -\vp_1-\vp_2)~ 
 \Gamma_{m}^{(2)}(-\vp_1, -\vp_2; R)   \nonumber \\ &&\times ~
 \Gamma_{m,~a}^{(3)}(\vk, \vp_1,\vp_2; R),\nonumber 
\end{eqnarray}
\begin{eqnarray}
\mK^{(1,~\rm 2lp)}_2  &=& 3~b_3(\vk+\vp_1+\vp_2,-\vp_1,-\vp_2)
\Gamma^{(3)}_{m,~a}(\vk, \vp_1, \vp_2;R) \nonumber \\
&& \times ~\Gamma_{m}^{(1)}(-\vp_1;R) \Gamma_{m}^{(1)}(-\vp_2;R),  
   \nonumber
\end{eqnarray}
\begin{eqnarray}
 \mK^{(1,~\rm 2lp)}_3  &=&4~ b_3(\vk-\vp_2,\vp_1+\vp_2,-\vp_1)~
 \Gamma_{m}^{(2)} (\vp_1, \vp_2;R) \nonumber \\
&& \times ~ \Gamma_{m}^{(1)}(-\vp_1;R) 
 \Gamma_{m,~a}^{(2)} (\vk,-\vp_2;R)  \nonumber
\end{eqnarray}
\begin{eqnarray}
 \mK^{(1,~\rm 2lp)}_4  &=& b_3(\vk, \vp_1+\vp_2, -\vp_1-\vp_2)~ 
\Gamma_{m}^{(2)}(\vp_1,\vp_2;R) \nonumber \\ &&\times~
\Gamma_{m}^{(2)}(-\vp_1,-\vp_2;R)
 \Gamma_{m,~a}^{(1)}(\vk;R)  \nonumber
\end{eqnarray}
\begin{eqnarray}
 \mK^{(1,~\rm 2lp)}_5  &=& b_4(\vk-\vp_1,\vp_2,-\vp_2,-\vp_1)~
  \left[ \Gamma_{m}^{(1)}(\vp_2;R) \right]^2 
\nonumber \\&& \times~
 \Gamma_{m}^{(1)}(\vp_1;R) 
 \Gamma_{m,~a}^{(2)}(\vk,-\vp_1;R)   \nonumber
\end{eqnarray}
\begin{eqnarray}
 \mK^{(1,~\rm 2lp)}_6  &=&b_4(\vk,\vp_1+\vp_2,-\vp_1,-\vp_2) ~
 \Gamma_{m}^{(2)}(\vp_1,\vp_2;R) \nonumber \\ && \times~  
 \Gamma_{m}^{(1)}(-\vp_1;R) 
 \Gamma_{m}^{(1)}(-\vp_2;R) 
 \Gamma_{m,~a}^{(1)}(\vk;R) 
 \nonumber
\end{eqnarray}
\begin{eqnarray}
 \mK^{(1,~\rm 2lp)}_7  &=& \frac{1}{8}~b_5(\vk, \vp_1,-\vp_1,\vp_2,-\vp_2)
 \left[ \Gamma_{m}^{(1)}(\vp_1;R) \right]^2 \nonumber \\ &&\times~
 \left[ \Gamma_{m}^{(1)}(\vp_2;R) \right]^2 
 \Gamma_{m,~a}^{(1)}(\vk;R) 
\end{eqnarray}

\appsection{One-loop Order of $\Gamma^{(2)}_{E}$}

\begin{figure*}[!htp]
\begin{center}
\includegraphics[width=0.75\textwidth]{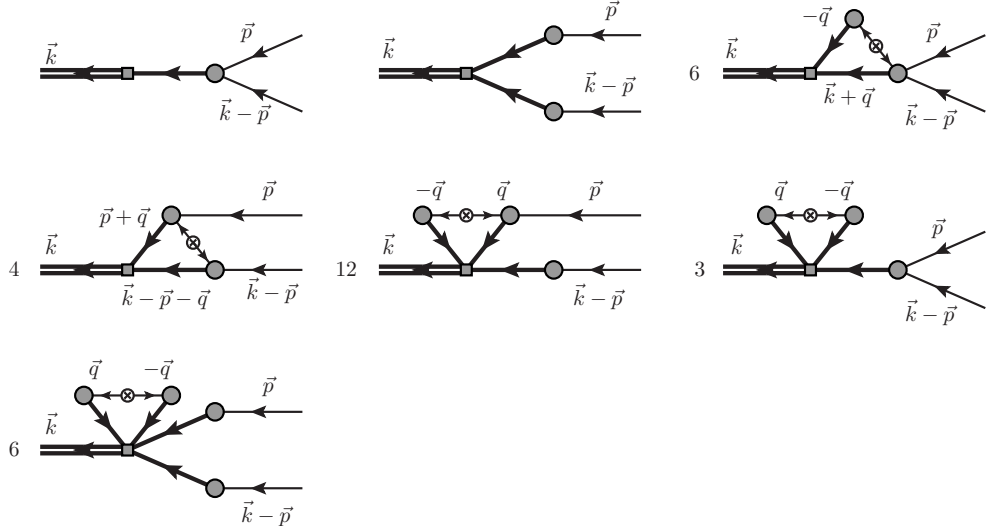}
\end{center}
\caption{\label{fig:GA2_1lp} Three-point nonlinear propagator
$\Gamma^{(2)}_{E}$ up to one-loop order.
}
\end{figure*}

At one-loop level, five diagrams are nonzero.
\begin{eqnarray}
\label{eqn:G2_1loop}
\Gamma^{(2,~{\rm 1lp})}_{E,~ab} (\vp,\vk-\vp; \eta) &= &
\int d^3\vp ~P_0(q) ~ \left[ 
 \sum_{i=1}^5 ~ \mK^{(\rm 2,~1lp)}_i \right]. \nonumber\\
\end{eqnarray}
where 
\begin{eqnarray}
\label{eqn:G2_1loop_ker}
\mK^{(\rm 2,~1lp)}_1 &=&3 ~b_2(\vk+\vq,-\vq)
\Gamma^{(3)}_{m,~ab}(\vk-\vp, \vp, \vq;R) \nonumber \\ && ~\times 
~\Gamma^{(1)}_{m}(-\vq;R) \nonumber
\end{eqnarray}

\begin{eqnarray}
\mK^{(\rm 2,~1lp)}_2 &=& 2 ~b_2(\vk-\vp-\vq,\vp+\vq)~  
~  \Gamma^{(2)}_{m,~a}(\vp,\vq;R)  \nonumber \\&&~\times
 \Gamma^{(2)}_{m,~b}(\vk-\vp,-\vq;R)  \nonumber 
 \end{eqnarray}

 \begin{eqnarray}
\mK^{(\rm 2,~1lp)}_3 &=& 2~b_3(\vk-\vp,\vp+\vq,-\vq)~
~  \Gamma^{(1)}_{m}(\vq;R) \nonumber\\ && ~\times
\Gamma^{(1)}_{m,~a}(\vk-\vp;R) 
 \Gamma^{(2)}_{m,~b}(\vp, \vq;R) \nonumber 
 \end{eqnarray}
 
 \begin{eqnarray}
\mK^{(\rm 2,~1lp)}_4 &=& \frac{1}{2}~ b_3(\vk, \vq, -\vq)~
 \Gamma^{(2)}_{m,~ab}(\vp, \vk-\vp;R) 
  \nonumber \\ && ~\times  \left[ \Gamma^{(1)}_{m}(\vq;R) \right]^2 \nonumber 
 \end{eqnarray}
  
\begin{eqnarray}
\mK^{(\rm 2,~1lp)}_5 &=& \frac{1}{4}~b_4(\vk-\vp, \vp, \vq, -\vq)~
      \Gamma^{(1)}_{m,~a}(\vk-\vq;R) \nonumber \\ && ~\times 
  ~ \left[\Gamma^{(1)}_{m}(\vq;R)\right]^2 
 \Gamma^{(1)}_{m,~b}(\vp;R).
\end{eqnarray}
In Eq.(\ref{eqn:G2_1loop_ker}), $\mK_1$ and $\mK_2$ are two-$\psid$ 
contributions from the first and second terms of Eq.(\ref{eqn:G2_n}) respectively,
$\mK_3$ and $\mK_4$ are three-$\psid$ contributions, and $\mK_5$ is the
four-$\psid$ contribution.

\label{lastpage}

\end{document}